\documentclass[aps,floatfix,showpacs,amsmath,amssymb]{revtex4}
\usepackage{natbib}
\usepackage{dcolumn}
\usepackage{graphicx}
\usepackage{color}
\usepackage[normalem]{ulem}
\usepackage{ulem}
\usepackage{epsfig}
\usepackage{float}
\newcommand{\apjl}{Astrophys.~J.~Lett.}

\newcommand{\phrd}{Phys.~Rev.~D.}
\newcommand{\jcap}{J.~Cosmol.~Astropart.~Phys.}
\newcommand{\be}{\begin{equation}}
\newcommand{\ee}{\end{equation}}
\newcommand{\bea}{\begin{eqnarray}}
\newcommand{\eea}{\end{eqnarray}}

\def\[{\begin{equation}}
\def\]{\end{equation}}
\begin{document}
\title{Expansion and Growth of Structure Observables in a Macroscopic Gravity Averaged Universe}
\author{Tharake Wijenayake\footnote{tsw091020@utdallas.edu}}
\author{Mustapha Ishak\footnote{mishak@utdallas.edu}}
\affiliation{
Department of Physics, The University of Texas at Dallas, Richardson, TX 75083, USA}
\date{\today}
\begin{abstract}
We investigate the effect of averaging inhomogeneities on expansion and large-scale structure growth observables using the exact and covariant framework of macroscopic gravity (MG). It is well-known that applying the Einstein's equations and spatial averaging do not commute and lead to the averaging problem and backreaction terms. For the MG formalism applied to the Friedman-Lemaitre-Robertson-Walker (FLRW) metric, the extra term can be encapsulated as an averaging density parameter denoted $\Omega_{\mathcal{A}}$. An exact isotropic cosmological solution of MG for the flat FLRW  metric is already known in the literature, we derive here an anisotropic exact solution. 
Using the isotropic solution, we compare the expansion history to current available data of distances to supernovae, Baryon Acoustic Oscillations, CMB last scattering surface data, and Hubble constant measurements, and find $-0.05 \le \Omega_{\mathcal{A}}  \le 0.07$ (at the 95\% confidence level). For the flat metric case this reduces to $-0.03 \le \Omega_{\mathcal{A}}  \le 0.05$. The positive part of the intervals can be rejected if a mathematical (and physical) prior is taken into account.
We also find that the inclusion of this term in the fits can shift the values of the usual cosmological parameters by a few to several percents. 
 Next, we derive an equation for the growth rate of large scale structure in MG that includes a term due to the averaging and assess its effect on the evolution of the growth compared to that of the $\Lambda$CDM concordance model. We find that an $\Omega_\mathcal{A}$ term of an amplitude range of [-0.04,-0.02] lead to a relative deviation of the growth from that of the $\Lambda$CDM of up to 2-4\% at late times.
Thus, the shift in the growth could be of comparable amplitude to that caused by similar changes in cosmological parameters like the dark energy density parameter or its equation of state. The effect could also be comparable in amplitude to some systematic effects considered for future surveys. This indicates that the averaging term and its possible effect need to be tightly constrained in future precision cosmological studies.
\end{abstract} 
\maketitle
%
\section{Introduction}\label{intro}
The rapid improvement in the quantity and quality of incoming and future observational data has encouraged the field of astrophysics to aim for a precise and accurate cosmology. In such a context, the understanding and control of systematic effects in the data and modeling have become essential to such an endeavor. Indeed, in order to constrain the information extracted from the data to a percent precision level, it is necessary to consider nuisance effects and contaminants that affect the data at this level. These effects include not only systematics related to data measurements but also corrections in the theoretical modeling including nonlinear and relativistic effects. 

One effect worth exploring is the averaging problem in relativity and cosmology that may effect the precision and accuracy of the cosmological constraints derived from the data \cite{Zalaletdinov:1992}-\cite{Zotov}. The problem originates from the fact that spatial averaging (or smoothing out inhomogeneities) in the universe is an operation that does not commute with applying the Einstein field equations. In other words, the field equations derived form the Friedmann-Lemaitre-Robertson-Walker (FLRW) metric that describes the universe at very large scales will be different from the equations derived at smaller scales and then averaged over large distances and volumes in the universe. This is due to the non-linear nature of the field equations of General Relativity. The presence of this non-commutation is well-known in the literature and usually gives additional terms in the Friedmann equations known as back-reaction terms \cite{Baumann:2010tm}-\cite{Wiltshire:2007a}. This back-reaction has been in general agreed upon by now to be too small to affect dramatically the overall dynamics of the universe, however it remains an open question whether this back-reaction can affect the constraints on cosmological parameters at the percent-precision level and thus if it should be considered at the same footing as other systematics in cosmological analyses. In this paper, we address some aspects of this question. 
 
The averaging procedures and the resulting backreaction terms provide mathematical formalisms on how smaller scale inhomogeneities in the universe can affect the dynamics at large scales. Such averaging schemes provide effective dynamical equations that explicitly relate the ``macroscopic" observables to the underlying ``microscopic" structure. The construction of averaging schemes for General Relativity has a long history and there has been many proposed methods \cite{Zalaletdinov:1992}-\cite{Zotov}. While some have proposed averaging schemes that use only scalar evolution equations, e.g. \cite{Futamase:1988,Futamase:1996,Kasai:1993,Buchert:2001,Buchert:2003,Coley:2009yz}, a great deal of effort was put into developing a fully covariant averaging procedure that can be applied to tensor equations \cite{Zalaletdinov:1992,Zalaletdinov:1993,Brannlund:2010rs}. 

Perhaps one of the most promising covariant schemes was proposed by \cite{Zalaletdinov:1992,Zalaletdinov:1993} and is referred to as Macroscopic Gravity in the literature. This was inspired by  some earlier work on the subject matter \cite{Isaacson:1968f,Isaacson:1968s}. 
The formalism derives the macroscopic gravitational field equations based on the usual Einstein's equation plus a tensor term due to the averaging process. 
These new field equations can be solved for a specific macroscopic geometry (for example the FLRW metric) without explicit reference to the microscopic geometries. The solutions to the macroscopic equations will give the dynamical equations with additional terms due to the averaging that are related to the microscopic geometries.
This became an attractive framework for cosmology and the Macroscopic Gravity field equations have been solved for the flat FLRW metric \cite{Coley:2005ei, vandenHoogen:2009nh,Clifton:2012fs}  and some authors have explored the cosmological implications of the formalism \cite{paranjape,Clarkson:2011gm}. 

In this paper we analyze the effect of terms due to the macroscopic gravity averaging scheme on constraints from the expansion history and the growth rate of large scale structure in the universe.  

The structure of the paper is as follows. In section \ref{AVG}, we give an overview of the Macroscopic Gravity formalism. Then, in section \ref{algo} we detail the approach to systematically obtain exact solutions in macroscopic gravity. We briefly rederive the flat FLRW isotropic solution and introduce a new anisotropic exact solution. In section \ref{exp}, we compare the macroscopic gravity observables of the isotropic solution to the expansion history data. In section \ref{instability} we derive the growth equation for the flat FLRW model. Finally we conclude in section \ref{conclusion}.
Units are chosen throughout the paper such that c = 1.
\section{The Averaging and Macroscopic Gravity Formalism}\label{AVG}

In this section we will introduce Zalaletdinov's  Macroscopic Gravity formalism detailed in Refs \cite{Zalaletdinov:1992,Zalaletdinov:1993}.
The Macroscopic Gravity formalism consists of a covariant averaging procedure, a method of assigning derivatives to the averaged geometric objects, and the application of the averaging procedure to the Einstein Field Equation (EFE) and the Cartan's Structure Equations in order to derive an effective EFE giving the coarse grained macroscopic dynamics.

The average of an arbitrary tensor field $P^{\alpha}_{\,\,\,\,\beta}$ at a point x over some averaging region  $ \Sigma_{x}$ surrounding the supporting point x is defined as
\be
\overline{P^{\alpha}_{\,\,\,\,\beta}}(x)  =\frac{1}{V_{\Sigma_{x}}} \int_{\Sigma_{x}} P^{\alpha'}_{\,\,\,\,\beta'}(x') \mathcal{A}^{\alpha}_{\,\,\,\,\alpha'}(x,x')\mathcal{A}^{\beta'}_{\,\,\,\, \beta}(x',x)\sqrt{-g(x')}d^4x'                     \label{avefield}
\ee

\noindent Where, $ V_{\Sigma_{x}} = \int_{\Sigma_{x}} \sqrt{-g(x')}d^4x'$ is the 4-volume of the averaging region for support point x, and $\mathcal{A}^{\alpha}_{\,\,\,\,\alpha'}(x,x')$, $\mathcal{A}^{\beta'}_{\,\,\,\, \beta}(x',x)$ are arbitrary bivectors (i.e. two point vectors whose primed index transforms like a vector at $x'$ and unprimed index transforms as a vector at x ) which satisfy the conditions  $\lim_{x'  \to x} \mathcal{A}^{\alpha}_{\,\,\,\,\beta'}(x',x) =\delta^{\alpha}_{\beta}$ and $\mathcal{A}^{\alpha}_{\,\,\,\,\beta'}(x,x')\mathcal{A}^{\beta'}_{\,\,\,\,\gamma''}(x',x'')=\mathcal{A}^{\alpha}_{\,\,\,\,\gamma''}(x,x'')$.
The first condition ensures that the average tensor at a supporting point x  ($\overline{P^{\alpha}_{\,\,\,\,\beta}}(x)$) becomes the value of the original tensor field at x when the averaging region goes to zero  ($ \lim_{\Sigma_{x} \to 0} \overline{P^{\alpha}_{\,\,\,\,\beta}}(x)   = P^{\alpha}_{\,\,\,\,\beta}(x)$) and the second ensures that $\mathcal{A}^{\alpha}_{\,\,\,\,\beta'}(x,x')$ is the inverse operator of $\mathcal{A}^{\alpha'}_{\,\,\,\,\beta}(x',x)$.
The most natural and well known bivector is what is known as the bivector of geodesic parallel displacement  ($g^{\alpha'}_{\,\,\,\,\beta}(x',x)$) \cite{{J.L.Synge:1960zz}} which satisfies both the required conditions and hence can be used as the ``averaging bivector". In fact, when the  averaging bivector is  the bivector of geodesic parallel displacement, the above definition of tensor averaging is similar to the one used by Isaacson in his well known paper on gravitational radiation \cite{Isaacson:1968f,Isaacson:1968s}  with the exception that in the latter, the integration is over the background while in the above definition, the integration is over the actual microscopic geometry, i.e. in the above equation the measure $\sqrt{-g(x')}$ is that of the microscopic metric rather than that of the macroscopic background metric. However, when defining the differentiation of average tensors, using the bivector of geodesic parallel displacement introduces some complications which we will discuss below, so it won't be the one used in the Macroscopic Gravity formalism. 

In order to define derivatives of average tensors, the averaging region at each  supporting point is defined according to the following prescription which Zalaletdinov calls ``averaging region coordination".
 All the points $x^{\alpha'}  \in \Sigma_{x}$ in a chosen averaging region ($\Sigma_{x}$) for a supporting point $x^{\alpha}$, are Lie dragged to a nearby supporting point $y^{\alpha} =x^{\alpha}+\Delta \lambda \xi^{\alpha}$ along the integral curves of an arbitrary vector field $\xi^{\alpha}$ parameterized by $\lambda$, in order to define the averaging region at that point ($\Sigma_{y}$),  using a bivector referred to as the coordination bivector which satisfies the two conditions satisfied by the previous averaging bivector. For simplicity, the averaging bivector can be taken as identical to the coordination bivector.   Now, the averaging region of supporting point y ($\Sigma_{y}$) read
 
 $ \Sigma_{y}   =\{y^{\alpha'}| y^{\alpha'} =  x^{\alpha'} +\Delta \lambda \xi^{\beta}\mathcal{A}^{\alpha'}_{\,\,\,\, \beta}(x',x) ;   x^{'\alpha} \in \Sigma_{x}\}$ 
 
 This procedure is used to construct averaging regions for all the supporting points in the manifold. The well defined coordination between the averaging regions allow us to write the measure in a region in terms of the measure in a nearby region as 
\be
 \sqrt{-g(y')} d^4y' = \sqrt{-g(x')} \left(1+\Delta \lambda \mathcal{A}^{\alpha'}_{\,\,\,\,\beta} \xi^{\beta} (\ln \sqrt{-g(x')})_{;{\alpha'} }    +\Delta \lambda (\mathcal{A}^{\alpha'}_{\,\,\,\,\beta} \xi^{\beta})_{;{\alpha'} }\right)+O({\Delta \lambda}^2)   \nonumber
\ee
Where the semicolon stands for covariant derivation with respect to the microscopic connection.\\
Using the above expression, the Lie derivative of the volume $V_{\Sigma_{x}}$ can be written as
 
 \be
 \pounds_{\bf{\xi}} V_{\Sigma_{x}} = \xi^{\alpha} \left<\mathcal{A}^{\beta'}_{\,\,\,\, \alpha \beta'} \right> V_{\Sigma_{x}}  \label{volpres}
\ee

\noindent  With the angle bracket denoting integration over the averaging region divided by $ V_{\Sigma}$, for example
 $\overline{P^{\alpha}} \equiv <\mathcal{A}^{\alpha}_{\;\;\alpha'}P^{\alpha'}>$.\\
\noindent The Lie derivative of an arbitrary average vector $\overline{P^{\alpha}}$ can be expressed as

\be
  \pounds_{\bf{\xi}} \overline{P^{\alpha}} = \xi^{\beta} \left( \left< \tilde{P^{\alpha}} _{:\beta}  \right>         +\left< \mathcal{A}^{\beta'}_{\,\,\,\, \beta ;  \beta'}P^{\alpha} \right>   -\left< \mathcal{A}^{\beta'}_{\,\,\,\, \beta ; \beta'} \right> \overline{P^{\alpha}}   \right)       \label{derave}
\ee
where we have defined ${P}_{: \alpha}    \equiv   {P}_{,\alpha} + \mathcal{A}^{\beta'}_{\,\,\,\, \alpha}{P}_{, \beta'}$, the coma stands for partial derivative, and the over tilde represents the bilocal extension of a geometric object, for example $\tilde{P^{\alpha}}  =\mathcal{A}^{\alpha}_{\;\;\alpha'}P^{\alpha'}$.

In order for the average tensors to be single valued functions of the supporting point, the partial derivatives must commute.
Since $ \pounds_{\xi} \overline{P^{\alpha}}  = \xi^{\beta} \left< P^{\alpha} \right>_{, \beta}  - \overline{P^{\beta}}\xi^{\alpha}_{\,\,\,\,  ,\beta}$    the commutator of the partial derivatives is given by
\be
\overline{P^{\alpha}}_{,[ \beta \gamma]}  = \left<   \tilde{P^{\alpha}}_{:[\beta \gamma]}\right> +  \left<  \tilde{P^{\alpha}} \mathcal{A}^{\delta}_{\,\,\,\, [ \gamma : \beta] ; \delta}  \right> - \left< \mathcal{A}^{\delta}_{\,\,\,\, [ \gamma : \beta] ; \delta}\right>\overline{P^{\alpha}}     \nonumber
\ee
Requiring the partial derivatives to commute ($\overline{P^{\alpha}}_{,[ \beta \gamma]}  = 0$) implies that the averaging bivector satisfies the condition
 \be
 \mathcal{A}^{\alpha'}_{\,\,\,\,[\beta, \gamma]} + \mathcal{A}^{\alpha'}_{\,\,\,\,[\beta, \delta'}\mathcal{A}^{\delta'}_{\,\,\,\, \gamma]}= 0  \label{cond1}
 \ee

\noindent Furthermore, from Equation (\ref{volpres})  the condition for the Lie-dragging of an averaging region to be volume preserving (i.e. $ \pounds_{\xi} V_{\Sigma_{x}} =0$) reads
\be
\mathcal{A}^{\alpha'}_{\,\,\,\, \beta ; \alpha'}   = 0 \label{cond2}
\ee

Now, using the conditions (\ref{cond1}),(\ref{cond2}) and equation (\ref{derave}), the partial derivatives of an average vector can be written as
\be
\overline{P^{\alpha }}_{,\beta} =\left<    \tilde{P^{\alpha}}_{:\beta} \right>
\ee

The bivector of parallel propagation does not in general satisfy conditions (\ref{cond1}) or  (\ref{cond2}), so it is not possible to set up the averaging region coordination using that as the averaging bivector. However, it has been shown \cite{Mars:1997jy} that for an arbitrary metric there always exists a bivector satisfying  equations (\ref{cond1}) and  (\ref{cond2}), and that satisfying these two conditions is equivalent to the bivector being the product of two vector bases $\mathcal{A}^{\alpha}_{\,\,\,\,\beta'}(x,x') = e^{\alpha}_{(i)}(x) e^{(i)}_{\beta'}(x') $ with the structure functions $C^{k}_{\,\,\,\ij}$ being constant (where $\left[  e_{(i)},e_{(j)} \right] = C^{k}_{\,\,\,\ij}e_{(k)}$). Choosing different $e_{i}$ (i.e. different averaging bivectors) will give different averaged fields for a given microscopic tensor field (see Eq. (\ref{avefield})).

In order to obtain an effective EFE, it is necessary to know what averaged geometric object gives the effective dynamics. In the Zalaletdinov formalism, the  ``bilocal extension of the connection coefficients" defined as
\be
{\mathcal{F}}^{\alpha}_{\,\,\,\, \beta \gamma} :=  \mathcal{A}^{\alpha}_{\,\,\,\, \epsilon'} \left(        \mathcal{A}^{ \epsilon'}_{\,\,\,\, \beta ,\gamma} + \mathcal{A}^{ \epsilon'}_{\,\,\,\, \beta ;\sigma'}\mathcal{A}^{\sigma'}_{\gamma}   \right)
\ee
which transforms like a connection at $x$, like a scalar at $x'$, and reduces to the microscopic connection $\Gamma^{\alpha}_{\,\,\,\, \beta \gamma}$ in the limit $x'$ goes to $x$, is what should be averaged in order to get the effective macroscopic connection coefficient.  There will be a macroscopic curvature tensor ($M^{\alpha}_{\,\,\,\, \beta \gamma \delta}$) and a macroscopic metric ($G_{\alpha \beta}$) corresponding to the macroscopic connection ($<{{\mathcal{F}}^{\alpha}_{\,\,\,\, \beta \gamma} }>$). Additionally, there exists a connection ($\pi^{\alpha}_{\,\,\,\, \beta \gamma}$) corresponding to the averaged microscopic Riemann tensor (${\bar{R}}^{\alpha}_{\,\,\,\, \beta \gamma \delta}$).
The difference between the two connection coefficients is defined as the  Affine deformation tensor  (${A}^{\alpha}_{\,\,\,\, \beta \gamma}= \,<{{\mathcal{F}}^{\alpha}_{\,\,\,\, \beta \gamma} }>-\pi^{\alpha}_{\,\,\,\, \beta \gamma}$).

By averaging out the Cartan structure equations, the metric compatibility equation , their integrability conditions and the microscopic EFE, the Macroscopic Gravity field equations can be constructed and shown to be of the form
\be
\overline{ g^{\alpha \epsilon}}M_{\alpha \gamma} -\frac{1}{2}\delta^{\epsilon}_{\gamma}\overline{ g^{\mu \nu}}M_{\mu \nu}  =8\pi G \left( \overline{ T^{\epsilon}_{\,\,\,\,\gamma}}  +{T^{(grav)}}^{\epsilon}_{\,\,\,\,\gamma}    \right)        \label{aveEFE}
 \ee
Where $M_{\beta \gamma}$ denotes the macroscopic Ricci tensor, $\overline{ g^{\alpha \epsilon} }$ the average of the inverse microscopic metric, $ \overline{ T^{\epsilon}_{\,\,\,\,\gamma}}$ the averaged stress energy tensor, and ${T^{(grav)}}^{\epsilon}_{\,\,\,\,\gamma}$ the gravitational stress energy tensor.
\be
8\pi G \,  {T^{(grav)}}^{\epsilon}_{\,\,\,\,\gamma}   = -\left(Z^{\epsilon}_{\,\,\,\, \mu \nu \gamma} + \frac{1}{2}\delta{\epsilon}{\gamma}Q_{\mu \nu}         \right)\overline{  g^{\mu \nu}}  \label{gravenergy}
 \ee

 Here the correlation 2-form $Z^{\alpha \,\,\,\, \,\,\,\, \mu}_{ \,\,\,\,\beta\gamma \,\,\,\,\nu\sigma}$ is defined as
 \be
Z^{\alpha \,\,\,\, \,\,\,\, \mu}_{ \,\,\,\,\beta\gamma \,\,\,\,\nu\sigma} =  \left<{\mathcal{F}}^{\alpha}_{\,\,\,\, \beta [\gamma}{\mathcal{F}}^{\mu}_{\,\,\,\, \underline{\nu} \sigma]}\right>  -\left<{\mathcal{F}}^{\alpha}_{\,\,\,\, \beta [\gamma}\right>\left<{\mathcal{F}}^{\mu}_{\,\,\,\, \underline{\nu} \sigma]}\right>   
 \ee
and it's traces are as follows.\\
$Q^{\alpha}_{\,\,\,\,\beta \rho \mu} = -2 Z^{\epsilon \,\,\,\, \,\,\,\, \alpha}_{ \,\,\,\,\beta\rho \,\,\,\,\epsilon\gamma}$,
  $Z^{\epsilon}_{\,\,\,\, \mu \nu \gamma}=2Z^{\epsilon \,\,\,\, \,\,\,\, \delta}_{ \,\,\,\,\mu\delta \,\,\,\,\nu\gamma}$ and $Q_{\mu \nu}  =Q^{\epsilon}_{\,\,\,\, \mu \nu \epsilon}=Z^{\delta}_{\,\,\,\, \mu \nu \delta}$ \\
where $Q^{\alpha}_{\,\,\,\,\beta \rho \mu}$ is known as the polarization tensor.\\

The correlation 2-form has the following symmetries 
 \be
 Z^{\alpha \,\,\,\, \,\,\,\, \mu}_{ \,\,\,\,\beta(\gamma \,\,\,\,\underline{\nu}\sigma)}=0   \label{sym1}
 \ee
 \be
 Z^{\alpha \,\,\,\, \,\,\,\, \mu}_{ \,\,\,\,\beta\gamma \,\,\,\,\nu\sigma} =-Z^{\mu \,\,\,\, \,\,\,\, \alpha}_{ \,\,\,\,\nu\gamma \,\,\,\,\beta\sigma}   \label{sym2}
 \ee
 \be
Z^{\alpha \,\,\,\, \,\,\,\, \mu}_{ \,\,\,\,\beta[\gamma \,\,\,\,\nu\sigma]}= 0   \label{cyclic} 
\ee
 and it satisfies the equi affine constraint
 \be
 Z^{\alpha \,\,\,\, \,\,\,\, \mu}_{ \,\,\,\,\alpha\gamma \,\,\,\,\nu\sigma} =0        \label{equiaffine}
 \ee
The differential properties for the correlation 2-form are set by a correlation 3-form and a correlation 4-form. It is possible to set the correlation 3 and 4-forms to zero and hence greatly simplify the formalism by setting the covariant derivative of the correlation 2-form with respect to the macroscopic connection to zero.
\be
Z^{\alpha \,\,\,\, \,\,\,\, \mu}_{ \,\,\,\,\beta[\gamma \,\,\,\,\underline{\nu}\sigma ||\lambda]}=0 \label{dz}
\ee
 where $||$ represents covariant derivative with respect to the macroscopic connection.
 This equation also ensures that the averaged stress energy tensor is conserved.
 
 The above equation has the integrability condition,
 \be
Z^{\epsilon \,\,\,\, \,\,\,\, \gamma}_{ \,\,\,\,\beta[\mu \,\,\,\,\underline{\delta}\nu}M^{\alpha}_{\,\,\,\, \underline{\epsilon} \kappa \pi]}   - Z^{\alpha \,\,\,\, \,\,\,\, \gamma}_{ \,\,\,\,\epsilon[\mu \,\,\,\,\underline{\delta}\nu}M^{\epsilon}_{\,\,\,\, \underline{\beta} \kappa \pi]}   +Z^{\alpha \,\,\,\, \,\,\,\, \epsilon}_{ \,\,\,\,\beta[\mu \,\,\,\,\underline{\delta}\nu}M^{\gamma}_{\,\,\,\, \underline{\epsilon} \kappa \pi]}   - Z^{\alpha \,\,\,\, \,\,\,\, \gamma}_{ \,\,\,\,\beta[\mu \,\,\,\,\underline{\epsilon}\nu}M^{\epsilon}_{\,\,\,\, \underline{\delta} \kappa \pi]}   = 0   \label{zm}
\ee
 
Furthermore, setting the correlation 3 and 4 -forms to zero require the quadratic constraint
 
\be
\begin{split}
Z^{\delta \,\,\,\, \,\,\,\, \theta}_{ \,\,\,\,\beta[\gamma \,\,\,\,\underline{\kappa}\pi}Z^{\alpha \,\,\,\, \,\,\,\, \mu}_{ \,\,\,\,\underline{\delta}\epsilon \,\,\,\,\underline{\nu}\sigma]}   +   Z^{\delta \,\,\,\, \,\,\,\, \mu}_{ \,\,\,\,\beta[\gamma \,\,\,\,\underline{\nu}\sigma}Z^{\theta \,\,\,\, \,\,\,\, \alpha}_{ \,\,\,\,\underline{\kappa}\pi \,\,\,\,\underline{\delta}\epsilon]}  +    Z^{\alpha \,\,\,\, \,\,\,\, \delta}_{ \,\,\,\,\beta[\gamma \,\,\,\,\underline{\nu}\sigma}Z^{\mu \,\,\,\, \,\,\,\, \theta}_{ \,\,\,\,\underline{\delta}\epsilon \,\,\,\,\underline{\kappa}\pi]}  
+  Z^{\alpha \,\,\,\, \,\,\,\, \mu}_{ \,\,\,\,\beta[\gamma \,\,\,\,\underline{\delta}\epsilon}Z^{\theta \,\,\,\, \,\,\,\, \delta}_{ \,\,\,\,\underline{\kappa}\pi \,\,\,\,\underline{\nu}\sigma]} \\ +  Z^{\alpha \,\,\,\, \,\,\,\, \theta}_{ \,\,\,\,\beta[\gamma \,\,\,\,\underline{\delta}\epsilon}Z^{\mu \,\,\,\, \,\,\,\, \delta}_{ \,\,\,\,\underline{\nu}\sigma \,\,\,\,\underline{\kappa}\pi]}  
  +Z^{\alpha \,\,\,\, \,\,\,\, \delta}_{ \,\,\,\,\beta[\gamma \,\,\,\,\underline{\kappa}\pi}Z^{\theta \,\,\,\, \,\,\,\, \mu}_{ \,\,\,\,\underline{\delta}\epsilon \,\,\,\,\underline{\nu}\sigma]}  =  0   \label{zz}
  \end{split}
\ee

The average of the Cartan equations implies the Affine deformation tensor needs to satisfy the constraint
 
 \be
 A^{\alpha}_{\,\,\,\, [\beta \sigma || \rho]} -  A^{\alpha}_{\,\,\,\, \epsilon [\rho} A^{\epsilon}_{\,\,\,\,\underline{\beta} \sigma]} =- \frac{1}{2}Q^{\alpha}_{\,\,\,\, \beta \rho\sigma}   \label{da}
 \ee
 
 Additionaly, the average of the integrability condition of the Cartan equations gives
 \be
 A^{\epsilon}_{\,\,\,\, \beta [\rho}M^{\alpha}_{\,\,\,\,\underline{\epsilon}\sigma\lambda]} +A^{\epsilon}_{\,\,\,\, \beta [\rho}Q^{\alpha}_{\,\,\,\,\underline{\epsilon}\sigma\lambda]} -    A^{\alpha}_{\,\,\,\, \epsilon [\rho}M^{\epsilon}_{\,\,\,\,\underline{\beta}\sigma\lambda]}   -    A^{\alpha}_{\,\,\,\, \epsilon [\rho}Q^{\epsilon}_{\,\,\,\,\underline{\beta}\sigma\lambda]}   = 0 \label{ar}
 \ee
  
For a given macroscopic metric,  equations ( \ref{sym1})-(\ref{ar}) can be solved to derive the correlation 2-form and the corresponding additional terms in the field equations. 

\section{Exact cosmological solutions to Macroscopic Gravity equations} \label{algo}

In order to solve the Macroscopic Gravity equations, it is essential to assume that the inverse of the averaged microscopic metric is equal to the macroscopic metric. This will restrict the class of solutions but it's not possible to avoid this since the theory does not provide a method of deriving this quantity other than explicitly performing the averaging.

\subsection{Algorithmic approach to solving the Macroscopic Gravity equations} 

A systematic approach to obtaining an exact solution are as follows.

\begin{itemize}
  \item Define the metric for the macroscopic  geometry  $G^{\alpha}_{\beta}$ and calculate the Riemannian curvature tensor $M^{\alpha}_{\,\,\,\, \beta \gamma \delta}$
  \item Define the correlation 2-form in terms of 720 arbitrary functions of the coordinates with the symmetries given by equation  (\ref{sym1})
  \item Apply the algebraic cyclic identity  Eq. (\ref{sym2}) 
  \item Apply the algebraic equi affine constraint  Eq. (\ref{equiaffine})
  \item  Solve the integrability condition Eq. (\ref{zm})
  \item  Solve the differential constraint  Eq. (\ref{dz})
  \item Solve the quadratic algebraic constraint for the correlation 2-form Eq. (\ref{zz})
\end{itemize}
Solving these equations will in general give all the independent components of the correlation 2 -form (although symmetries in the macroscopic geometry can place further constraints reducing the number of independent components).
\begin{itemize}
  \item Now the affine deformation tensor can be solved for using equations (\ref{da}) and (\ref{ar})
  \item The gravitational stress energy tensor can now be calculate using Eq. (\ref{gravenergy})
  \item Finally any constraints on the gravitational stress energy tensor due to symmetries in the macroscopic geometry need to be applied
\end{itemize}
All remaining independent functions in the correlation 2-form and the affine deformation tensor will correspond to different microscopic geometries giving the same macroscopic geometry, and are free parameters of the model.  Now the macroscopic EFE (\ref{aveEFE}) can be derived  for a given averaged stress energy tensor.

\subsection{Previously derived spatially homogeneous and isotropic solutions} 

The model of cosmological interest is the  one with the macroscopic geometry described by the FLRW metric in agreement with observations. The macroscopic gravity solutions for the flat FLRW metric has been studied in the  literature \cite{Coley:2005ei,vandenHoogen:2009nh,Clifton:2012fs} with the second reference giving a systematic analysis of the solutions for the case when the correlation 2-form  ($Z^{\alpha \,\,\,\, \,\,\,\, \mu}_{ \,\,\,\,\beta\gamma \,\,\,\,\nu\sigma}$) and the affine deformation tensor (${A}^{\alpha}_{\,\,\,\, \beta \gamma}$) are invariant under the six parameter group of Killing vectors (corresponding to the three translational and three rotational symmetries of the metric), and the electric part of the correlation tensor is zero. The solution to the correlation 2-form was found to be completely specified by three arbitrary constants $\mathcal{A}$, $h_2$ and $b_1$ while the affine deformation tensor was specified by only $\mathcal{A}$. The gravitational stress energy tensor reads:
\begin{equation}
8\pi G{T^{(grav)}}^{   \alpha}_{ \beta} = \left( \begin{array}{cccc}
\frac{\mathcal{A}^2}{a^2} & 0 & 0&0 \\
0& \frac{1}{3}\frac{\mathcal{A}^2}{a^2}& 0&0 \\
0& 0&\frac{1}{3} \frac{\mathcal{A}^2}{a^2}&0 \\
0 &0 & 0&\frac{1}{3} \frac{\mathcal{A}^2}{a^2} \end{array} \right)
\end{equation}
We have re-derived this solution in this work and our results are in agreement with the findings of \cite{Coley:2005ei,vandenHoogen:2009nh}. In summary, for the macroscopic line element 
\be
ds^2 = -dt^2+a^2 \left(dx^2 +dy^2 + dz^2 \right)  \label{feqn}
\ee
with an averaged stress energy tensor of the form of a perfect fluid  $\bar{T}^{\alpha}_{\beta} =diag (-\rho ,p,p,p)$ where $\rho $ is the energy density and p is the anisotropic pressure.\\
The macroscopic EFE  [\ref{aveEFE}] read
\be
\frac{\dot{a}^2}{a^2} = \frac{8\pi G}{3}\rho -\frac{1}{3}\frac{\mathcal{A}^2}{a^2} +\frac{\Lambda}{3}
\ee
\be
\frac{2\ddot{a}}{a}+\frac{\dot{a}^2}{a^2}  = -8 \pi Gp - \frac{1}{3}\frac{\mathcal{A}^2}{a^2} +\Lambda
\ee
\noindent where over-dots denote partial differentiation with respect to the time coordinate t.\\
Hence, the Macroscopic Gravity correlations appear like an extra positive spatial curvature term in the Friedmann's equations.

The constant $\mathcal{A}$ emerges from applying the formalism to the macroscopic flat FLRW metric. It has no explicit scale dependence, however, it does implicitly depend on scale in the sense that the derived expression holds only when the averaging is performed at a scale large enough to reduce the macroscopic geometry to completely homogeneous and isotropic. At smaller scales, the effects due to averaging will not be captured by this simple expression and will presumably explicitly depend on the scale. 

\subsection{Spatially homogeneous and anisotropic solutions to Macroscopic Gravity equations}  \label{apend}

The exact solutions for Macroscopic Gravity are known only for the flat homogeneous and isotropic and the static spherically symmetric cases \cite{Coley:2005ei,vandenHoogen:2009nh,Clifton:2012fs, VanDenHoogen:2007en}. The non static spherically symmetric solution has been found \cite{Coley:2006}  using ``volume preserving coordinates" and an approximation rather than by solving  equations (\ref{sym1})-(\ref{ar}) directly. 
In this section we will consider the solution for a macroscopically homogeneous, anisotropic and spatially flat metric  (i.e a macroscopically Bianchi type I metric) of the form
\be
dS^2 = -dt^2 + a(t)^2dx^2 +b(t)^2dy^2 +c(t)^2dz^2.
\ee

We note that we will derive this exact solution here just as a further example for macroscopic gravity but we will use for the observables and the remaining of the paper the isotropic solution from the previous sub-section.  

 We will not assume that correlation 2-form $Z^{\alpha \,\,\,\, \,\,\,\, \mu}_{ \,\,\,\,\beta\gamma \,\,\,\,\nu\sigma}$ is invariant under the three parameter group of Killing vectors ($G_3$). However,  the gravitational stress energy tensor (Eq. (\ref{gravenergy})) will be required to be diagonal and invariant under the action of $G_3$ since the average stress energy tensor being considered is  invariant under the action of $G_3$. In the language of \cite{Clifton:2012fs} it's a ``Type II" solution. 

The assumptions used to obtain the solutions are as follows,

\begin{itemize}
  \item The average of the inverse  microscopic metric is equal to inverse macroscopic metric \\
   $\bar{g}^{\alpha \beta}$ $=$ $G^{\alpha \beta}$
  \item The averaged microscopic stress energy tensor takes the form \\
  $\bar{T}^{\alpha}_{\,\,\,\,\beta}$ = diag[$-\rho (t)$,$p_1(t)$,$p_2(t)$,$p_3(t)$]
  \item The electric part of the correlation 2-form is zero.\\
   $Z^{\alpha \,\,\,\, \,\,\,\, \mu}_{ \,\,\,\,\beta\gamma \,\,\,\,\nu\sigma} u^{\sigma}    = 0$ where $u^{\sigma}=[1,0,0,0]$ is the time like vector orthogonal to the hyper-surface of homogeneity
   \item The affine deformation tensor will be assumed to be invariant under the action of the group of Killing vectors.
   $ \pounds_{\bf{k_{(i)}}}  {A}^{\alpha}_{\,\,\,\, \beta \gamma} = 0$  where  $\bf{k_{(i)}} = \partial_i$ and i=x,y,z

\end{itemize}

All the following calculations were performed using the publicly available tensor algebra package GRTensor and the commercial computer algebra package Maple.\\
From equation (\ref{sym1}) the correlation 2-form ostensibly has 720 independent component. We started by defining the correlation 2-form with these symmetries in terms of 720 functions of the coordinates. The cyclic identity (\ref{cyclic}) gives 250 independent constraints while the Equi-affine relation (\ref{equiaffine}) gives 76 additional independent linear constraints and the assumption  that the electric part of the correlation tensor is zero gives a further 275, reducing the number of independent components to 121. The calculations up-to this point will be true for any metric since the metric and it's derivatives played no role in the equations. So for any macroscopic geometry, the correlation 2-form will have at most 121 independent components.\\
Now applying the integrability condition (\ref{zm}) gives an additional 52 constraints bringing the total number of independent components to 69. Solving the differential constraint (\ref{dz}) forces all the functions to be independent of time and we are left with 69 functions of the position coordinates.\\
The gravitational stress energy tensor can now be calculated and it will be a diagonal and depend on 6 of the functions.  Applying the requirement that the gravitational stress energy tensor is invariant under the action of $G_3$ gives 6 differential constraints. Solving them, the gravitational stress energy tensor reads
\begin{equation}
8\pi G{T^{(grav)}}^{\alpha}_{\,\,\,\,\beta} = \frac{1}{a(t)b(t)c(t)} \left( \begin{array}{cccc}
\mathcal{A} a+\mathcal{B} b+\mathcal{C} c & 0 & 0&0 \\
0& \mathcal{A} a& 0&0 \\
0& 0&\mathcal{B} b&0 \\
0 &0 & 0&\mathcal{C} c \end{array} \right)
\end{equation}

where $\mathcal{A}, \mathcal{B}, \mathcal{C}$ are constants.
The correlation 2-form now has 69 independent components, 66 of them are functions of the position coordinates and 3 are constants.

Now the macroscopic EFE (\ref{aveEFE}) reads

\be
\frac{\dot{a}\dot{b}}{ab} +\frac{\dot{a}\dot{c}}{ac} +\frac{\dot{b\dot{c}}}{ac}  = -\frac{\mathcal{A} a+\mathcal{B} b+\mathcal{C} c}{abc} + 8\pi G \rho   \label{anefe1}
\ee
\be
\frac{\ddot{b}}{b}+\frac{\ddot{c}}{c}+\frac{\dot{b}\dot{c}}{bc} = -\frac{\mathcal{A}}{bc} -8\pi G p_1   \label{anefe2}
\ee

\be
\frac{\ddot{a}}{a}+\frac{\ddot{c}}{c}+\frac{\dot{a}\dot{c}}{ac} = -\frac{\mathcal{B}}{ac} -8\pi G p_2   \label{anefe3}
\ee

\be
\frac{\ddot{a}}{a}+\frac{\ddot{b}}{b}+\frac{\dot{b}\dot{c}}{ab} = -\frac{\mathcal{C}}{ab} -8\pi G p_3   \label{anefe4}
\ee

It would be of interest to know how this compares to the dynamics of a homogeneous anisotropic space time with spatial curvature. 
Since the RW metric with positive curvature is a special case of the Bianchi type IX, the natural choice for comparison would be of that type.

The metric for a Bianchi model can in general be written in the form
\be
dS^2 = -dt^2 + \eta_{(i) (j)} {w}^{(i)}_{\,\,a}{w}^{(j)}_{\,\,b} dx^a dx^b
\ee

where ${w}^{(i)}_{\,\,a}$ are the components of the invariant basis 1-forms corresponding to the Bianchi type and $\eta_{(i) (j)}$ is a symmetric matrix that is a function of only time.
For the Bianchi IX model the invariant basis 1-forms are  \cite{mix} 
\be
{w}^{(1)}  = \cos(\psi)d\theta +\sin(\psi)\sin(\theta)d\phi \nonumber
\ee
\be
{w}^{(2)} = \sin(\psi)d\theta \cos(\psi)\sin(\theta)d\phi  \nonumber
\ee
\be
{w}^{(3)} =d\psi +i \cos(\theta) d\phi   \nonumber
\ee

The simplest Bianchi IX model for comparison would be the one with $\eta_{(i) (j)} = diag[a(t),b(t),c(t)]$. For this model the EFE read

\be
\frac{\dot{a}\dot{b}}{ab} +\frac{\dot{a}\dot{c}}{ac} +\frac{\dot{b\dot{c}}}{ac}  = - \frac{2a^2b^2+2b^2c^2+2a^2c^2-a^4 -b^4 -c^4 }{4a^2b^2c^2} +  8\pi G \rho    \label{bx1}
\ee
\be
\frac{\ddot{b}}{b}+\frac{\ddot{c}}{c}+\frac{\dot{b}\dot{c}}{bc} =-\frac{2a^2b^2+2a^2c^2 +b^4+c^4 -3a^4 -2b^2c^2}{4a^2b^2c^2} -8\pi G p_1  \label{bx2}
\ee

\be
\frac{\ddot{a}}{a}+\frac{\ddot{c}}{c}+\frac{\dot{a}\dot{c}}{ac} =-\frac{2a^2b^2+2b^2c^2 +a^4+c^4 -3b^4 -2a^2c^2}{4a^2b^2c^2} -8\pi G p_2  \label{bx3}
\ee

\be
\frac{\ddot{a}}{a}+\frac{\ddot{b}}{b}+\frac{\dot{b}\dot{c}}{ab} =-\frac{2a^2c^2+2b^2c^2 +a^4+b^4 -3c^4 -2a^2b^2}{4a^2b^2c^2} -8\pi G p_3  \label{bx4}
\ee

The exact Macroscopic Gravity solutions for the spatially flat anisotropic metric (\ref{anefe1}-\ref{anefe4}) have some similarity with the spatially closed anisotropic solution for the microscopic EFE. When $a(t)=b(t)=c(t)$, both  sets of equations have the same form. Equations (\ref{anefe1}-\ref{anefe4}) reduce to the flat homogeneous solution described in the previous section while equations (\ref{bx1}-\ref{bx4}) reduce to the spatially closed FLRW solution. 
However, unlike the simple case of the isotropic solution, it is unclear how to relate these new terms to a spatial curvature.
It is known that the ``mixmaster" models described by equations (\ref{bx1})-(\ref{bx4}) show chaotic behavior at early times \cite{Misner:1967zz} and it remains to be analyzed  whether the macroscopically anisotropic models have similar behavior that would wipe out any macroscopic anisotropies.

\section{Expansion history observables and constraints on macroscopic gravity isotropic solution}\label{exp}

The observational consequence of the macroscopic FLRW model with  its additional ``dynamical curvature"  $\mathcal{A}^2/a^2$, on the luminosity distance measurements and hence the constraints on the cosmological parameters from the distance observables has been studied in the literature  \cite{Clarkson:2011gm}. In this paper we will apply the additional constraint $\mathcal{A}^2 \ge 0$ (hence $\Omega_\mathcal{A} \le 0$) coming from the constraint equations for the affine deformation tensor (see \cite{vandenHoogen:2009nh}). We obtain the results for the cases with and without this constraint.\\

First, we write the Macroscopic RW metric as, 
\begin{equation}
dS^{2} =-dt^{2}+ {a(t)}^{2} [dr^{2} +{f_{k}(r)^{2}}({d\theta}^{2}   +\sin^{2}\theta {d\phi}^{2})]  
\end{equation}

\begin{align}  \text{where} &&
  f_{k}(r)=\begin{cases}
   \sin(r) & \text{if } k = 1\,, \\ \nonumber
   r & \text{if }k =0\,, \\
   \sinh(r) & \text{if }k \leq 0\,.
  \end{cases}       && \text{and k is the spatial curvature of the macroscopic metric.}
\end{align}

 If the source is a perfect fluid, the effective EFE read:
 \be
H^2 = \frac{8}{3} \pi G \rho -\frac{k}{a^2} +\frac{\Lambda}{3}-\frac{1}{3}\frac{\mathcal{A}^2}{a^2} \label{fried}
\ee

\be
\frac{\ddot{a}}{a} = \frac{4}{3} \pi G\left(  \rho  +3p  \right) +\frac{1}{3}\Lambda
\ee

\noindent where the Hubble parameter has been defined as, $H=\dot{a}/a$.
The effective Friedman equation (\ref{fried}) can be rewritten in terms of the current matter parameters as

\begin{equation}
H(a) = H_{0}{ (\Omega_{k}a^{-2} +  \Omega_{\mathcal{A}}a^{-2}   +\Omega_{\Lambda}+\Omega_{m} a^{-3} )}^{\frac{1}{2}}   \label{hubble}
\end{equation}

\noindent where $\Omega_m\equiv\frac{8}{3} \pi G \rho /H_{0}^2$ is matter density parameter, $\Omega_{\Lambda}\equiv\Lambda /3H_{0}^2$ is the cosmological constant density parameter, $\Omega_{k}\equiv -k /a^{2}H_{0}^2$ is the curvature density parameter,  $\Omega_{\mathcal{A}}= -\mathcal{A}^2/{3 H_{0}^2 a^2}$ is the ``gravitational energy" parameter due to averaging \cite{Clarkson:2011gm}, and $H_{0}$ is the Hubble parameter evaluated today.\\

{In this work, we make the assumption that light rays on average follow the null geodesics of the averaged macroscopic space time and that the only changes to the luminosity distance are due to the change in the modified Friedmann equation. Some rays of light are  demagnified and some are magnified but on average photon flux conservation leads to no net change in the luminosity distance \cite{Weinberg}. So overall, this seems to be a reasonable assumption for the  average of a large number of photons  and is consistent with some of the findings in the literature \cite{Bull:2012zx, 2012JCAP...02..036D,1976ApJ...208L...1W}. However some authors have found \cite{2014JCAP...10..073B, Rasanen:2008jcai,2012PhRvD..85h3528R} that  inhomogeneities could lead to small changes in the redshift relation and possibly large changes to the luminosity distance. It is re-assuring that the changes in the redshift were found to be small in those studies while the changes in the luminosity distance are suppressed on average by photon flux conservation. Nevertheless, it will be good to address this point further in the formalism used in this paper by applying it to the null geodesic equation and the Sachs equations explicitly, similar to what was done for the EFE. We leave this full project for future work.}

Now, following the usual derivation, the luminosity distance can be written for the curved FLRW macroscopic metric as,
\begin{equation}
d_{L}  = \frac{1}{ a H_0 \sqrt{|\Omega_{k}|} } f_{k}\left( \int^{a}_{a'=1}  \frac{ \sqrt{  |\Omega_{k}|}da'}{{ (\Omega_{k}{a'}^{2} +\Omega_{\mathcal{A}}{a'}^{2}     +\Omega_{\Lambda} {a'}^4+\Omega_{m} {a'} )}^{\frac{1}{2}}
}\right)
\label{curved_mg_dl}
\end{equation}
and for the flat FLRW macroscopic metric it reads  
\begin{equation}
d_{L}  = \frac{1}{a H_0} \int^{a}_{a'=1} \frac{da'}{{(\Omega_{\mathcal{A}}{a'}^{2}     +\Omega_{\Lambda} {a'}^4+\Omega_{m} {a'} )}^{\frac{1}{2}}}.
\label{flat_mg_dl}
\end{equation}
Due the degeneracy  between $\Omega_\mathcal{A}$ and $\Omega_k$ in the denominator (i.e. Friedmann equation), for the fits to the data we use the dynamical energy term that is the sum of the averaging gravitational energy and geometric curvature ($\Omega_{kd}=\Omega_\mathcal{A} +\Omega_k$) same as ref \cite{Clarkson:2011gm}.

We fit the cosmological parameters for the FLRW solution of Macroscopic Gravity (and other models) using the available cosmological distance data. The supernova (SNe) observations from the Union 2.1 data set \cite{Suzuki:2011hu}, Cosmic Microwave Background (CMB) last scattering surface data from WMAP 9 year data release \cite{Hinshaw:2012aka}, Baryonic Acoustic Oscillations (BAO) from WiggleZ \cite{Blake:2011en}  and the Hubble rate from HST measurements \cite{Riess:2009pu}. 

The parameter fits for the various models were performed using $\chi^2$ minimization via a maximum likelihood analysis (i.e. we minimize $\chi^2 =\chi^2_{SN} +\chi^2_{BAO} +\chi^2_{CMB}) $ and Monte-Carlo Markov Chain approach using a modified version of the publicly available package COSMOMC \cite{Lewis:2002ah}. \\

In order to get the constraints from the supernova data, we define $\chi_{SN}^2$ as
\be
\chi^2_{SN} = \sum_{1=1}^{557} \frac{\left( \mu_{obs} (z_i)- \mu(z_i) \right)^2}{\sigma_i^2}
\ee
where $\mu(z) = \tilde{m}-M=5\log_{10}\left( d_L(z)\right) +25$ is the extinction corrected distance modulus, $\sigma_i$ is the uncertainty in the $i^{th}$ SNe data point, $\tilde{m}$ is the apparent luminosity and $d_L$ is the Luminosity distance measured in Mpc. When performing the fits we effectively marginalize over the absolute luminosity M.\\
In order to fit for the CMB surface of last scattering we define  three fitting parameters \cite{Komatsu:2010fb}.  The shift parameter R defined by
\be
R(z_*) =\sqrt{\Omega_m}(1+z_*)D_A(z_*)
\ee
the redshift to the surface of last scattering $z_*$ given by
\be
z_* = 1048\left(  1 +0.00124(\Omega_b h^2)^{-0.738}       \right)\left( 1+g_1(\Omega_mh^2)^{g_2} \right)
\ee
where  (see for example \cite{Hu:1995en} )
\be
 g_1 = \frac{0.0783(\Omega_bh^2)^{-0.238}}{1+39.5(\Omega_bh^2)^{0.763}}  \nonumber
\ee
\be
g_2= \frac{0.560}{1+21.1(\Omega_bh^2)^{1.81}} \nonumber
\ee
and  the acoustic scale ($l_a$)  defined as
\be
l_a =(1+z_*)\frac{\pi D_A(z_*)}{r_s(z_*)}
\ee
with the proper angular diameter distance, $D_A(z) =d_L(z)/(1+z)^2      $ and the comoving sound horizon
\be
r_s(z_*)=\frac{1}{\sqrt{3}} \int_{0}^{\frac{1}{1+z_*}}   \frac{da}{a^2H(a)\sqrt{1+(3\Omega_b /4\Omega_\gamma)a}}   
\ee
where $\Omega_{\gamma}  =2.469*10^{-5}h^{-2}  $ for   $T_{CMB}  =2.725K$.\\
The parameters are fitted using
\be
\chi_{CMB}^2 =\Delta x_i Cov^{-1}(x_i,x_j) \Delta x_j
\ee
with $x_i = (R,l_a,z_*)$, $\Delta x_i =x_i -x_i^{obs}$ and $Cov^{-1}(x_i,x_j)$ the inverse covariance matrix for the parameters from the reference \cite{Komatsu:2010fb}.

Next, in order to obtain the constraints from the BAO, following \cite{Eisenstein:2005su} we define the effective distance $D_V $ as
\be
D_v(z) =\left(  D_a^2(z)(1+z)^2 \frac{z}{H(z)} \right)^{\frac{1}{3}}
\ee
with the redshift at the decoupling epoch given by
\be
z_d =\frac{1291(\Omega_mh^2)^{0.251}}{1+0.659(\Omega_mh^2)^{0.828}}\left(1+b_1(\Omega_bh^2)  \right)
\ee
with
\be
b_1 = 0.313(\Omega_mh^2)^{-0.419}\left(  1+0.607(\Omega_mh^2)^{0.674}\right)  \nonumber
\ee
\be
b_2=0.238(\Omega_mh^2)^{0.233}  \nonumber
\ee

The parameter constraints from the BAO are now given by
\be
\chi^2_{BAO} = \sum_{i}  \left(\frac{r_s(z_i)/D_v(z_i)-(r_s(z_i)/D_v(z_i))_{obs}}{\sigma_i}\right)^2
\ee

We also add the prior, the inverse of the angular diameter distance at red shift 0.04  equals  $6.49405 \times10^{-3} \pm 0.31512\times10^{-3}$, (i.e. $H_0  =74.2 \pm 3.6$ km/s for the fiducial model) given by the HST measurements \cite{Riess:2009pu},  and the prior $\Omega_bh^2 =0.022 \pm 0.002$ given by big bang nucleosynthesis.  

\begin{figure}

\begin{center}

\begin{tabular}{c}

{\includegraphics[width=5.0in,height=4.0in,angle=0]{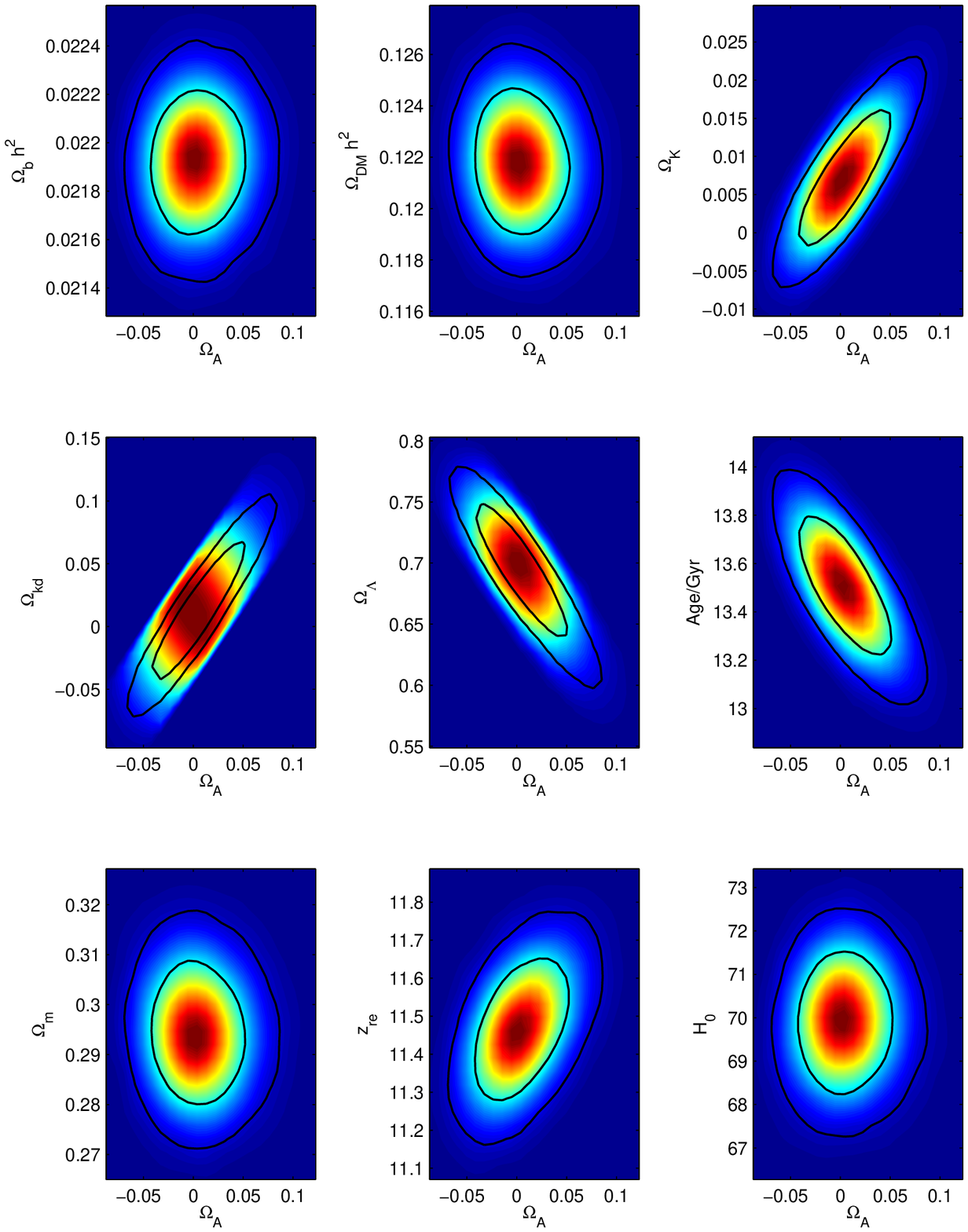}}\\
\\
\\
\end{tabular}

\caption{\label{fig1}
Two dimensional marginalized joint contour plots ((68\% and 95\% confidence levels)) for the FLRW solution of the Macroscopic Gravity using the 
supernova data from the Union2.1 compilation, the HST data, the last scattering surface data from WMAP9, and the WiggleZ BAO data. These are the results for the Macroscopic Gravity model without any prior on $\Omega_\mathcal{A}$ and where spatial curvature is also allowed. }
\end{center}

\end{figure}

We perform the parameter fits for Macroscopic Gravity, $\Lambda$CDM and wCDM with a constant equation of state, by varying the physical dark matter density  ($\Omega_{DM}h^2$), the physical baryon density ($\Omega_{b}h^2$),  the curvature parameter ($\Omega_k$) and in the cases of Macroscopic Gravity and wCDM, the dynamic curvature i.e. the sum of the averaging gravitational energy and geometric curvature ($\Omega_{kd}=\Omega_\mathcal{A} +\Omega_k$) and the equation of state of dark energy (w) respectively. The values for those parameters and the derived parameters $\Omega_m,\Omega_\mathcal{A}, \Omega_{\Lambda} $ and $H_0$ are summarized in Table \ref{table1}, and Figure \ref{fig1}.

For the macroscopic gravity, we find in  the case of models restricted by the mathematical and physical prior \cite{vandenHoogen:2009nh}, that $-0.027 \le \Omega_A \le 0$ (68\% confidence level). In the case where we do not impose the prior, we obtain $-0.024 \le \Omega_A \le 0.036$.  As we will discuss further in section \ref{conclusion}, the mathematical prior turned out also to be a physical prior consistent with the fact that a larger magnitude of negative backreaction term leads to a larger enhancement of the growth of structure as supported by studies using inhomogeneous cosmological models \cite{Peel:2012vg, Bolejko}. We are also able to reproduce exactly the results of reference \cite{Clarkson:2011gm} where the table shows that the constrained value for $\Omega_A$ is significantly large  when the SDSS SNe1A compilation  \cite{Kessler:2009ys} (which uses the MLCS2k2 light curve fitter)  is used.
Table \ref{table1}, and Figure \ref{fig1} uses only SNe data from the Union 2.1 compilation that uses the SALT II light curve fitter.    

\begin{table}[h] 
\centering  
\begin{tabular}{ccccccc} 
\hline\hline  \\[0.5ex]    

Parameters &$ \Lambda$CDM& wCDM&  MG with prior& MG w/o prior&  Flat MG with prior&Flat MG w/0 prior \\ [0.5ex]  
\hline\hline    	
\\ 
$\Omega_K$&$0.014^{+0.018}_{-0.018}$&$-0.013^{+0.017}_{-0.017}$&$0.0026^{+0.0047}_{-0.0047}$&$0.0075^{+0.0059}_{-0.0057}$& 0&0 \\   
\\
$\Omega_\Lambda$&$0.688^{+0.041}_{-0.040}$&$0.690^{+0.040}_{-0.040}$&$0.724^{+0.020}_{-0.019}$&$0.692^{+0.036}_{-0.036}$&$0.733^{+0.016}_{-0.015}$&$0.711^{+0.023}_{-0.023}$	\\  
\\
$\Omega_m$&$0.298^{+0.025}_{-0.025}$&$0.297^{+0.025}_{-0.025}$&$0.295^{+0.014}_{-0.014}$&$0.295^{+0.091}_{-0.092}$&$0.279^{+0.013}_{-0.013}$&$0.280^{+0.013}_{-0.013}$\\  
\\
$H_0$&$71.5^{+2.7}_{-2.7}$&$72.8^{+3.0}_{-3.0}$&$69.9^{+1.6}_{-1.6}$&$69.9^{+1.0}_{-1.0}$&$69.5^{+1.2}_{-1.2}$&$72.8^{+3.0}_{-3.0}$\\   
\\
$\Omega_\mathcal{A}$&{N/A}&{N/A}&$-0.0216^{+0.0216}_{-0.0054}$&$0.0058^{+0.0299}_{-0.0299}$& $-0.0123^{+0.0123}_{-0.0098}$& $0.009^{+0.019}_{-0.019}$\\  
\\
w&{-1}&$-1.12^{+0.11}_{-0.11}$&{-1}&{-1}&{-1}&{-1}
\\
\hline
\end{tabular} 
\caption{Marginalized parameter constraints (68 \% confidence) from the cosmological distance observations (supernova data from the Union2.1 compilation, the HST data, the last scattering surface data from WMAP9, and the WiggleZ BAO data). The prior here is $\Omega_{\mathcal{A}} \le 0.$. Results are given for the Macroscopic Gravity using the curved and flat macroscopic FLRW metric, with and without the prior.} 
\label{table1} 
\end{table}

\section{Growth of Large Scale Structure in the macroscopic gravity averaged universe}\label{instability}
\subsection{Derivation of the growth evolution equations in Macroscopic Gravity}

In order to study the growth of structure due to small inhomogeneities in the macroscopically  flat Friedmann universe, we will perturb the metric, the stress energy tensor, the correlation 2-form and the affine deformation tensor about the exact macroscopic solution. The new quantities will be given by
\be
       g^{\alpha}_{\,\,\,\, \beta} = {g^{(0)}}^{\alpha}_{\,\,\,\,\beta} +\delta{ g^{\alpha}_{\,\,\,\,\beta}}
\ee
\be
       \bar{T}^{\alpha}_{\,\,\,\, \beta} =  {\bar{T}}^{(0) \alpha}_{\,\,\,\,\,\,\,\,\,\,\,\,\beta} +\delta{ \bar{T}^{\alpha}_{\,\,\,\,\beta}}
\ee
\be
      Z^{\alpha \,\,\,\, \,\,\,\, \mu}_{ \,\,\,\,\beta\gamma \,\,\,\,\nu\sigma} =  {Z^{(0)}}^{\alpha \,\,\,\, \,\,\,\, \mu}_{ \,\,\,\,\beta\gamma \,\,\,\,\nu\sigma} +\delta{ Z^{\alpha \,\,\,\, \,\,\,\, \mu}_{ \,\,\,\,\beta\gamma \,\,\,\,\nu\sigma}}
\ee
\be
      {A}^{\alpha}_{\,\,\,\, \beta \gamma} =  {A^{(0)}}^{\alpha}_{\,\,\,\, \beta \gamma}+\delta{ {A}^{\alpha}_{\,\,\,\, \beta \gamma}}
\ee
where the superscript zero denotes the value of the exact solution and the prefix $\delta$ denotes the perturbations about the exact value. The perturbations are taken to be of order $\epsilon$ . By expanding the macroscopic gravity equations (\ref{aveEFE})-(\ref{ar})  in terms of the order parameter up to linear order, it's possible to obtain the equations governing the first order terms.  The perturbations we consider will be on scales smaller than the averaging domain,  so they will be the fluctuations that would be smoothed out from the averaging.

We reproduced the perturbation results of \cite{Clifton:2012fs}, but present here the derivation steps, and we derive the growth rate equation further below. 

The zeroth order terms will satisfy the original exact equations while the first order terms will satisfy a linearized version of the above equations (see \cite{Wald:1984rg}).
The equations (\ref{sym1})-(\ref{ar}) at first order will be given by
 \be
\delta Z^{\alpha \,\,\,\, \,\,\,\, \mu}_{ \,\,\,\,\beta(\gamma \,\,\,\,\underline{\nu}\sigma)}=0   \label{sym1first}
 \ee
 \be
 \delta Z^{\alpha \,\,\,\, \,\,\,\, \mu}_{ \,\,\,\,\beta\gamma \,\,\,\,\nu\sigma} =-\delta Z^{\mu \,\,\,\, \,\,\,\, \alpha}_{ \,\,\,\,\nu\gamma \,\,\,\,\beta\sigma}   \label{sym2first}
 \ee
 \be
\delta Z^{\alpha \,\,\,\, \,\,\,\, \mu}_{ \,\,\,\,\beta[\gamma \,\,\,\,\nu\sigma]}= 0   \label{cyclicfirst} 
\ee
 \be
 \delta Z^{\alpha \,\,\,\, \,\,\,\, \mu}_{ \,\,\,\,\alpha\gamma \,\,\,\,\nu\sigma} =0        \label{equiaffinefirst}
 \ee

\be
\delta Z^{\alpha \,\,\,\, \,\,\,\, \mu}_{ \,\,\,\,\beta[\gamma \,\,\,\,\underline{\nu}\sigma |\lambda]}=0 \label{dzfirst}
\ee
 \be
 \begin{split}
\delta Z^{\epsilon \,\,\,\, \,\,\,\, \gamma}_{ \,\,\,\,\beta[\mu \,\,\,\,\underline{\delta}\nu}M^{\alpha}_{\,\,\,\, \underline{\epsilon} \kappa \pi]}   - \delta Z^{\alpha \,\,\,\, \,\,\,\, \gamma}_{ \,\,\,\,\epsilon[\mu \,\,\,\,\underline{\delta}\nu}M^{\epsilon}_{\,\,\,\, \underline{\beta} \kappa \pi]}   +\delta Z^{\alpha \,\,\,\, \,\,\,\, \epsilon}_{ \,\,\,\,\beta[\mu \,\,\,\,\underline{\delta}\nu}M^{\gamma}_{\,\,\,\, \underline{\epsilon} \kappa \pi]}   - \delta Z^{\alpha \,\,\,\, \,\,\,\, \gamma}_{ \,\,\,\,\beta[\mu \,\,\,\,\underline{\epsilon}\nu}M^{\epsilon}_{\,\,\,\, \underline{\delta} \kappa \pi]}   \\
{Z^{(0)}}^{\epsilon \,\,\,\, \,\,\,\, \gamma}_{ \,\,\,\,\beta[\mu \,\,\,\,\underline{\delta}\nu}\delta M^{\alpha}_{\,\,\,\, \underline{\epsilon} \kappa \pi]}   -  {Z^{(0)}}^{\alpha \,\,\,\, \,\,\,\, \gamma}_{ \,\,\,\,\epsilon[\mu \,\,\,\,\underline{\delta}\nu}\delta M^{\epsilon}_{\,\,\,\, \underline{\beta} \kappa \pi]}   + {Z^{(0)}}^{\alpha \,\,\,\, \,\,\,\, \epsilon}_{ \,\,\,\,\beta[\mu \,\,\,\,\underline{\delta}\nu}\delta M^{\gamma}_{\,\,\,\, \underline{\epsilon} \kappa \pi]}   - {Z^{(0)}}^{\alpha \,\,\,\, \,\,\,\, \gamma}_{ \,\,\,\,\beta[\mu \,\,\,\,\underline{\epsilon}\nu}\delta M^{\epsilon}_{\,\,\,\, \underline{\delta} \kappa \pi]}   = 0   \label{zmfirst}
 \end{split}
\ee
\be
\begin{split}
\delta Z^{\delta \,\,\,\, \,\,\,\, \theta}_{ \,\,\,\,\beta[\gamma \,\,\,\,\underline{\kappa}\pi}{Z^{(0)}}^{\alpha \,\,\,\, \,\,\,\, \mu}_{ \,\,\,\,\underline{\delta}\epsilon \,\,\,\,\underline{\nu}\sigma]}   +  \delta Z^{\delta \,\,\,\, \,\,\,\, \mu}_{ \,\,\,\,\beta[\gamma \,\,\,\,\underline{\nu}\sigma}{Z^{(0)}}^{\theta \,\,\,\, \,\,\,\, \alpha}_{ \,\,\,\,\underline{\kappa}\pi \,\,\,\,\underline{\delta}\epsilon]}  +    \delta Z^{\alpha \,\,\,\, \,\,\,\, \delta}_{ \,\,\,\,\beta[\gamma \,\,\,\,\underline{\nu}\sigma}{Z^{(0)}}^{\mu \,\,\,\, \,\,\,\, \theta}_{ \,\,\,\,\underline{\delta}\epsilon \,\,\,\,\underline{\kappa}\pi]}  
+   \delta Z^{\alpha \,\,\,\, \,\,\,\, \mu}_{ \,\,\,\,\beta[\gamma \,\,\,\,\underline{\delta}\epsilon}{Z^{(0)}}^{\theta \,\,\,\, \,\,\,\, \delta}_{ \,\,\,\,\underline{\kappa}\pi \,\,\,\,\underline{\nu}\sigma]} \\ +   \delta Z^{\alpha \,\,\,\, \,\,\,\, \theta}_{ \,\,\,\,\beta[\gamma \,\,\,\,\underline{\delta}\epsilon}{Z^{(0)}}^{\mu \,\,\,\, \,\,\,\, \delta}_{ \,\,\,\,\underline{\nu}\sigma \,\,\,\,\underline{\kappa}\pi]}  
  + \delta Z^{\alpha \,\,\,\, \,\,\,\, \delta}_{ \,\,\,\,\beta[\gamma \,\,\,\,\underline{\kappa}\pi}{Z^{(0)}}^{\theta \,\,\,\, \,\,\,\, \mu}_{ \,\,\,\,\underline{\delta}\epsilon \,\,\,\,\underline{\nu}\sigma]} 
  +{Z^{(0)}}^{\delta \,\,\,\, \,\,\,\, \theta}_{ \,\,\,\,\beta[\gamma \,\,\,\,\underline{\kappa}\pi} \delta Z^{\alpha \,\,\,\, \,\,\,\, \mu}_{ \,\,\,\,\underline{\delta}\epsilon \,\,\,\,\underline{\nu}\sigma]}   +   {Z^{(0)}}^{\delta \,\,\,\, \,\,\,\, \mu}_{ \,\,\,\,\beta[\gamma \,\,\,\,\underline{\nu}\sigma} \delta Z^{\theta \,\,\,\, \,\,\,\, \alpha}_{ \,\,\,\,\underline{\kappa}\pi \,\,\,\,\underline{\delta}\epsilon]}  \\ +    {Z^{(0)}}^{\alpha \,\,\,\, \,\,\,\, \delta}_{ \,\,\,\,\beta[\gamma \,\,\,\,\underline{\nu}\sigma} \delta Z^{\mu \,\,\,\, \,\,\,\, \theta}_{ \,\,\,\,\underline{\delta}\epsilon \,\,\,\,\underline{\kappa}\pi]}  
+  {Z^{(0)}}^{\alpha \,\,\,\, \,\,\,\, \mu}_{ \,\,\,\,\beta[\gamma \,\,\,\,\underline{\delta}\epsilon} \delta Z^{\theta \,\,\,\, \,\,\,\, \delta}_{ \,\,\,\,\underline{\kappa}\pi \,\,\,\,\underline{\nu}\sigma]}  +  {Z^{(0)}}^{\alpha \,\,\,\, \,\,\,\, \theta}_{ \,\,\,\,\beta[\gamma \,\,\,\,\underline{\delta}\epsilon} \delta Z^{\mu \,\,\,\, \,\,\,\, \delta}_{ \,\,\,\,\underline{\nu}\sigma \,\,\,\,\underline{\kappa}\pi]}  
  +{Z^{(0)}}^{\alpha \,\,\,\, \,\,\,\, \delta}_{ \,\,\,\,\beta[\gamma \,\,\,\,\underline{\kappa}\pi} \delta Z^{\theta \,\,\,\, \,\,\,\, \mu}_{ \,\,\,\,\underline{\delta}\epsilon \,\,\,\,\underline{\nu}\sigma]}=  0   \label{zzfirst}
 \end{split}
\ee
 \be
 \delta A^{\alpha}_{\,\,\,\, [\beta \sigma | \rho]} -   \delta A^{\alpha}_{\,\,\,\, \epsilon [\rho} {A^{(0)}}^{\epsilon}_{\,\,\,\,\underline{\beta} \sigma]}-{A^{(0)}}^{\alpha}_{\,\,\,\, \epsilon [\rho} \delta A^{\epsilon}_{\,\,\,\,\underline{\beta} \sigma]} =- \frac{1}{2} \delta Q^{\alpha}_{\,\,\,\, \beta \rho\sigma}   \label{dafirst}
 \ee
 \be
 \begin{split}
 \delta A^{\epsilon}_{\,\,\,\, \beta [\rho}M^{\alpha}_{\,\,\,\,\underline{\epsilon}\sigma\lambda]} + \delta A^{\epsilon}_{\,\,\,\, \beta [\rho}Q^{\alpha}_{\,\,\,\,\underline{\epsilon}\sigma\lambda]} -     \delta A^{\alpha}_{\,\,\,\, \epsilon [\rho}M^{\epsilon}_{\,\,\,\,\underline{\beta}\sigma\lambda]}   -     \delta A^{\alpha}_{\,\,\,\, \epsilon [\rho}Q^{\epsilon}_{\,\,\,\,\underline{\beta}\sigma\lambda]}   \\
  {A^{(0)}}^{\epsilon}_{\,\,\,\, \beta [\rho} \delta M^{\alpha}_{\,\,\,\,\underline{\epsilon}\sigma\lambda]} +  {A^{(0)}}^{\epsilon}_{\,\,\,\, \beta [\rho}\delta Q^{\alpha}_{\,\,\,\,\underline{\epsilon}\sigma\lambda]} -     {A^{(0)}}^{\alpha}_{\,\,\,\, \epsilon [\rho}\delta M^{\epsilon}_{\,\,\,\,\underline{\beta}\sigma\lambda]}   -     {A^{(0)}}^{\alpha}_{\,\,\,\, \epsilon [\rho}\delta Q^{\epsilon}_{\,\,\,\,\underline{\beta}\sigma\lambda]}= 0 \label{arfirst}
   \end{split}
 \ee
 where $|$ represents covariant derivative with respect to the unperturbed macroscopic connection.

If  $ {Z^{(0)}}^{\alpha \,\,\,\, \,\,\,\, \mu}_{ \,\,\,\,\beta\gamma \,\,\,\,\nu\sigma}, {A^{(0)}}^{\epsilon}_{\,\,\,\, \beta \rho}  \sim O(\epsilon)$ (i.e. ${\mathcal{A}}^2 ,h_1,b_2 \sim O(\epsilon) $) the equations governing the first order correlation 2-form and the affine deformation tensor will be identical to the ones satisfied by the zeroth order quantities (\ref{sym1})-(\ref{ar}). Therefore if we assume that electric part of the first order correlation tensor is zero ($z^{\alpha \,\,\,\, \,\,\,\, \mu}_{ \,\,\,\,\beta\gamma \,\,\,\,\nu\sigma}v^{\sigma} = 0$ where $v^{\sigma} = \frac{1}{a}(1-\phi,\delta u^i )$ ) the first order gravitational stress energy tensor will have the form of a positive spatial curvature term.\\
The assumption that ${Z^{(0)}}^{\alpha \,\,\,\, \,\,\,\, \mu}_{ \,\,\,\,\beta\gamma \,\,\,\,\nu\sigma} \sim O(\epsilon)$ is consistent with the small mean value for $\Omega_{\mathcal{A}}$ obtained in section \ref{exp}.

Since only scalar perturbations are relevant for the growth of inhomogeneities we will restrict the metric perturbations to just the  scalar part. Now, without any loss of generality the metric can be written in the conformal Newtonian gauge as
\begin{equation}
dS^2 =a(\eta)^{2} (-(1+2\phi)d\eta^{2}  +(1-2\psi)(dx^2 +dy^2 +dz^2))
\end{equation}

The source will be considered as a perturbed perfect fluid and the first order stress energy tensor will read 
\begin{align}
\delta{ \bar{T}^{\eta}_{\,\,\,\,\eta}} = -\delta{\rho}  \\ \delta{ \bar{T}^{\eta}_{\,\,\,\,i}}  =\frac{1}{a}(\rho +p)\delta{u_i} \\ \delta{ \bar{T}^{i}_{\,\,\,\,j}}=\delta p \,\,{\delta}^i_j
\end{align}

where $\delta\rho$ is the energy density perturbation,  $\delta p$ is the pressure perturbation and $\delta{u_i} $ is the comoving peculiar velocity

Now the modified Einstein field equation (\ref{aveEFE}) at first order read
\begin{eqnarray}
\nabla^2  \phi -3\mathcal{H}(\mathcal{H}\phi +\phi')  =4 \pi G a^2( \delta \rho+ {\delta \rho}_{\mathcal{A}})  \label{efe1}\\
\nabla_i {(\mathcal{H}\phi +\phi')} =-4 \pi G a^2\left(p+\rho-\frac{2\mathcal{A}^2}{3a^2}\right)\nabla_i{\delta u} \label{efe2}\\ 
\phi'' +3\mathcal{H}\phi' +(2{\mathcal{H}}'+{\mathcal{H}}^2 )\phi =4\pi G a^{2} \left(\delta p -\frac{{\delta \rho}_{\mathcal{A}}}{3}\right) \label{efe3}
\end{eqnarray}
where $\nabla_{i}$ is the spatial covariant derivative, ${\delta \rho}_{\mathcal{A}}$ is the energy perturbation to the  gravitational stress energy tensor, a prime denotes the derivative with respect to $\eta$, $\nabla_i \delta{u}$ is the irrotational part of the comoving peculiar velocity of the fluid (which can be written as a divergence of a function  $\delta{u}$)   and $\mathcal{H}$ is defined as $\frac{a'}{a}$. Equation (\ref{efe1}) is the (0,0) component, equation (\ref{efe2}) is the (i,0) component and equation (\ref{efe3})  is the (i,j) component  where i$\neq$ j  . When there is no anisotropic stress the  i = j component gives $\psi=\phi$  and has been used to eliminate $\psi$ from the above equations.

At this point it is convenient to decompose the perturbations into the eigenfunctions of the Laplace equation (see \cite{kodamaetall1984}).
Scalar harmonics satisfy     $\nabla^2 Q+k^2 Q=0 $   while vector harmonics are given by    $ Q_{i} = \frac{\nabla_{i} Q}{k}$ . 

It becomes apparent that for subhorizon modes  (k$\eta$ $>>$1) $\phi$ and  $\phi '$ are negligible compared to the spatial derivatives of $\phi$. Hence  equation (\ref{efe1}) can be rewritten as
\begin{equation}
\nabla ^2 \phi  =4 \pi G  a^2 \rho\left(\delta  +\frac{{\delta \rho}_{\mathcal{A}} }{\rho} \right)          \label{poisson}
\end{equation}

The twice contracted Bianchi identity for the modified EFE  gives
\begin{equation}
\bar{T}^{\alpha}_{\,\,\beta||\alpha} +{T}^{(grav) \alpha}_{\,\,\,\,\,\,\,\,\,\,\,\,\,\,\,\,\,\,\,\,\beta||\alpha} = 0
\end{equation}
It might at first appear that the stress energy components are not independently conserved. However that is not the case.
It can be shown that the differential constraint on the correlation 2 form (\ref{dz}) implies the gravitational stress energy tensor is conserved  (${T^{(grav)}}^{\alpha}_{\,\,\,\,\beta||\alpha} = 0$) and hence, the averaged stress energy tensor is conserved independently.

The first order conservation equations for the averaged stress energy tensor are given by
 \begin{eqnarray}
 \delta \rho ' +3\mathcal{H}(\delta p +\delta \rho )-3 \phi '(\rho +p) +a(p+\rho)\nabla ^2\delta u=0  \label{ec}  \\
\frac{1}{a^4} {((\rho+p)a^5\nabla ^2\delta u)}'    +\nabla^2\delta p +(\rho +p)\nabla^2 \phi=0   \label{mc}
\end{eqnarray}

Where, the first equation comes from $\delta \bar{T}^{\alpha}_{\,\,\,\,0|\alpha}=0$ and the second equation comes from the spatial divergence of $ \delta \bar{T}^{\alpha}_{\,\,\,\,i|\alpha} =0$.
Since we are considering the matter dominated era, the radiation can be neglected.
Hence, p = 0, and $(\rho a^3)$ is a constant. 
Defining the density contrast by $\delta_{m}\equiv \delta\rho/\rho$ and using equations (\ref{ec}) and (\ref{mc}), we can obtain an obtain an evolution equation for the density contrast of the form:
\begin{equation}
\delta_m^{''} +\mathcal{H}\delta_m^{'} -4\pi G a^2 \rho\left(\delta_m  +\frac{\delta \rho_{\mathcal{A}}}{\rho}\right) =0  \label{denev}
\end{equation}

In order to proceed, it is necessary to write the perturbation to the gravitational energy density  in terms of the matter energy density. In order to do that we argue that even though the matter stress energy tensor and the gravitational stress energy tensor are conserved independently, the perturbation to the gravitational energy density must be tightly coupled to the perturbation of the matter energy density, since the inhomogeneities in the matter cause the gravitational stress energy. Hence, we can assume that the comoving peculiar velocity of the gravitational energy density is the same as the matter comoving peculiar velocity.
With this assumption, the ${T^{(grav)}}^{\alpha}_{\,\,\,\,0||\alpha} = 0$  component of the first order conservation equations of the gravitational stress energy tensor is given by
\begin{equation}
(\delta_{\mathcal{A}}  -2 \phi) ' +\frac{2}{3}a \nabla \delta u =0
\label{gravcons} 
\end{equation}
where 
\begin{equation}
\delta_{\mathcal{A}} \equiv\frac{\delta{\rho}_{\mathcal{A}}}{\left[\frac{1}{8 \pi G}\frac{\mathcal{A}^2}{a^2}\right]}   
\end{equation}

Using equations (\ref{gravcons}) and (\ref{denev}) to eliminate $\delta u $ and integrating,  since $\delta_{\mathcal{A}} = 0$ when $\delta_m= 0$, we find:
\begin{equation}
\delta_m -\frac{3}{2}\delta_{\mathcal{A}} =0   \nonumber
\end{equation}
Which gives
\begin{equation}
{\delta \rho}_{\mathcal{A}} = -\frac{\mathcal{A}^2 \delta_m}{12\pi G a^2}
\end{equation}
Substituting the above in equation (\ref{denev}) gives the growth equation
\begin{equation}
\delta_m^{''} +\mathcal{H}\delta_m^{'} -\left(4\pi G a^2\rho  -\frac{\mathcal{A}^2}{3a^2}\right)\delta_m =0  
\end{equation}
The above equation in terms of the cosmological time t, reads:
\begin{equation}
\ddot{\delta_m} +2H\dot{\delta_m} -\left(4\pi G \rho -\frac{\mathcal{A}^2}{3a^2 }\right)\delta_m =0  \nonumber
\end{equation}
Where the dot denotes differentiation with respect to t, and H is the Hubble parameter $\dot{a}/a$.

In order to conveniently plot the growth,  the growth equation can be written as a function of scale factor
\begin{equation}
\delta_m^{''} +\left((\ln H)^{' }  +\frac{3}{a}\right) \delta_m^{'} -  \frac{1}{a^2 H^2}\left(  4\pi G \rho -\frac{\mathcal{A}^2}{3a^2 }\right)\delta_m =0  \label{growth}
\end{equation}
where  prime now denotes partial differentiation with respect to a rather than $\eta$. 
\begin{figure}[t]
\centering
  \fbox{\includegraphics[width=.65\linewidth]{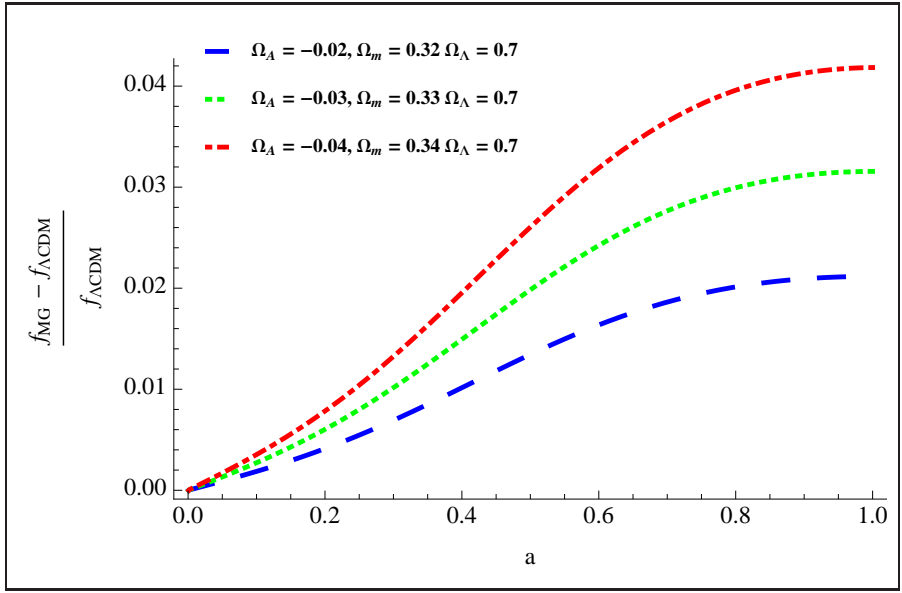}}
\caption{The relative difference in growth factor between the Macroscopic Gravity model MG-$\Lambda$CDM and the $\Lambda$CDM. A non-zero contribution of the $\Omega_\mathcal{A}$ backreaction term due to averaging increases the growth proportionally to the amplitude of this term. The effect of $\Omega_\mathcal{A}$ is up to 2-4\% at late times on the growth factor function compared to the $\Lambda$CDM. }
\label{fig2}
\end{figure}

\subsection{Effect of averaging on the growth rate versus precision cosmology requirements}

In order to compare the growth of structure within a given model to observational data, it is most common to use the logarithmic growth factor since that is what is measured from for example redshift distortions and Lyman-Alpha forests \cite{19,20,21,22,23,24,25,26,27}
\begin{equation}
f = \frac{d \, \ln\, \delta}{d \,  \ln \,a}
\end{equation}
It would be of interest to know how the effects of back-reaction on the growth compares to that of the dark energy density of equation of state. Particularly whether the effects of averaging can be degenerate with a change in these parameters. For dynamical dark energy models with a constant equation of state w (see for example  \cite{dossett, linder,polarski,Gong09,Mortonson,linder07,gannouji,ygong,PaperI}), the growth equation is given by 
\be
\frac{d\,f}{d\, \ln{a}} +f^2 +\left( \frac{\dot{H}}{H^2} +2 \right)f -\frac{3}{2} \Omega_{m} a^{-3} \frac{H_{0}^{2}}{H^2} =  0
\ee
with
\be
\frac{{H_{0}}^2}{H^2} =  \Omega_{m} a^{-3} + \Omega_{K} a^{-2} + \Omega_{\Lambda}  a^{-3(1+w)} \nonumber
\ee
\be
\frac{\dot{H}}{H^2} =\frac{1}{2}\frac{H_{0}^2}{H^2}\left( -3\Omega_{m} a^{-4}  -2\Omega_{K} a^{-3} -3(1+w)\Omega_{\Lambda}a^{-3(1+w)}  \right)  \nonumber
\ee
Substituting $\delta^{'} =\frac{\delta}{a}f$ and $\delta^{''}=\frac{\delta}{a^2}(f^2 -f +\frac{d\, f}{d \,ln \,a})$ in equation (\ref{growth}) we obtain the growth equation for the Macroscopic Gravity model in terms of the growth factor as
\be
\frac{d\,f}{d\,\ln {a}} +f^2 +\left( \frac{\dot{H}}{H^2} +2 \right)f -\left(\frac{3}{2} \Omega_{m} a^{-3}   + \Omega_{\mathcal{A}}a^{-2}\right)\frac{H_{0}^{2}}{H^2} =  0
\ee
where
\be
\frac{H^{2}}{H_{0}^{2}}   =  \Omega_{m} a^{-3}  +\Omega_{K} a^{-2} +\Omega_{\mathcal{A}} a^{-2}+ \Omega_{\Lambda}^{0}  \nonumber
\ee
\be
\frac{\dot{H}}{H^2} =\frac{1}{2}\frac{H_{0}^2}{H^2}\left( -3\Omega_{m} a^{-4}  -2\Omega_{K} a^{-3} -2\Omega_{\mathcal{A}}a^{-3}  \right)  \nonumber
\ee

\begin{figure}
\begin{center}
\begin{tabular}{|c|c|}
\hline
{\includegraphics[width=3.3in,height=2.5in,angle=0]{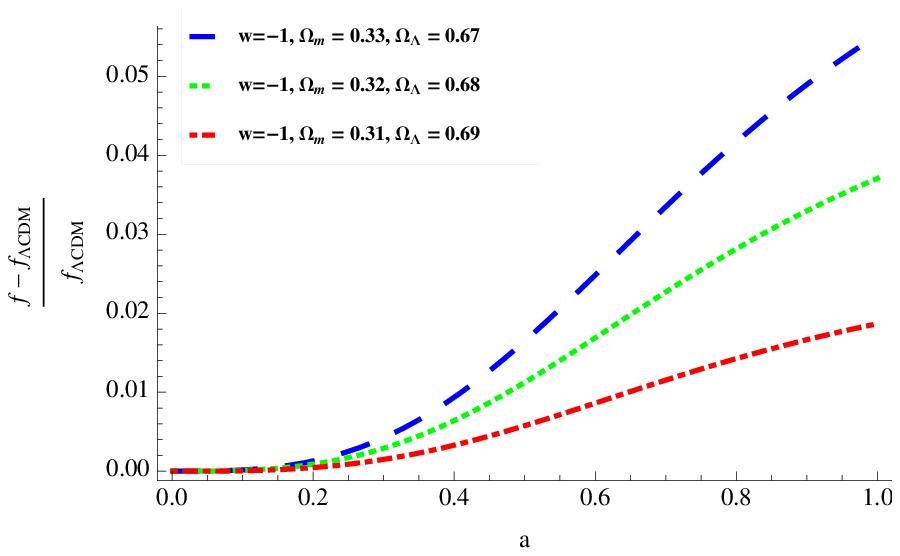}} &
{\includegraphics[width=3.3in,height=2.5in,angle=0]{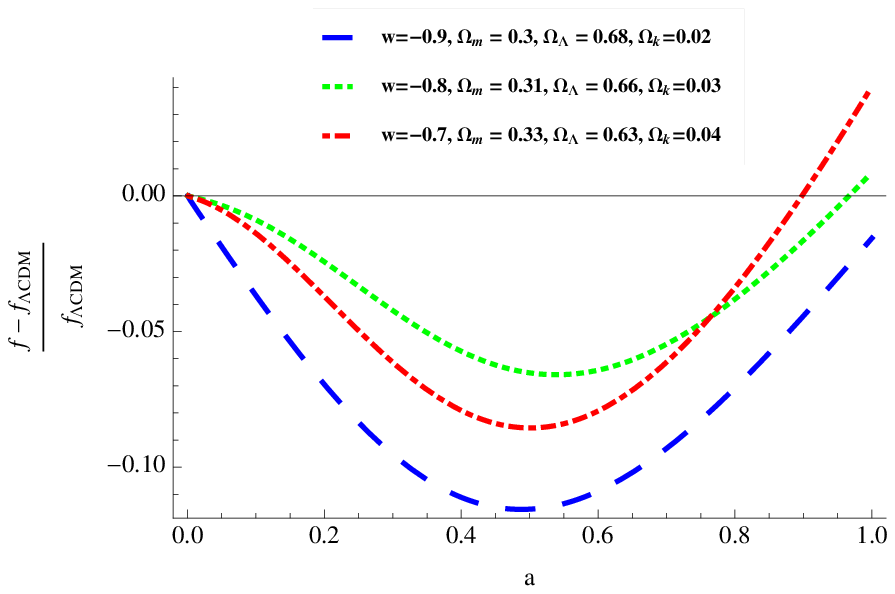}}\\
\hline
\end{tabular}
\caption{\label{fig3}
LEFT:  The relative difference in the growth factor function between the $\Lambda$CDM model with $\Omega_{\Lambda}=0.70$ and those where $\Omega_{\Lambda}$ takes the values shown on the figure. Increasing $\Omega_{\Lambda}$ increases the late-time growth suppression as expected. 
RIGHT: The relative difference of the growth factor between the $\Lambda$CDM model (i.e. the equation of state $w=-1$) and dark energy models where the equation of state takes the values shown on the figure. In both cases, the growth factor function is changed by up to several percent.}
\end{center}
\end{figure}
We find that a non-zero negative $\Omega_\mathcal{A}$ term of $2-4\%$ due to averaging has the effect of enhancing the growth by $2-4\%$ at late times relative to when no averaging backreaction is taken into account (see Fig. \ref{fig2}). These effects are of the same order as those resulting from changing the dark energy density parameter or its equation of state (see Fig. \ref{fig3}).  

\section{Conclusion}\label{conclusion}
%
In this work we studied the effects of averaging inhomogeneities on the expansion history and the growth rate of large scale structure using the non perturbative framework of Macroscopic Gravity. The framework is based an exact mathematical formalism developed to provide a covariant averaging procedure. The formalism results in modified Friedmann equations with a new term that can be viewed as a back-reaction term and have been previously called as the averaging gravitational energy density parameter $\Omega_A$.  
 
As examples of exact solutions to macroscopic gravity field equations, we rederive here a previous isotropic and homogeneous solution and we obtain a new homogeneous but anisotropic solution. Starting from the macroscopically homogeneous, anisotropic, and spatially flat metric of Bianchi type-I, we derive the effective Einstein field equations with new terms due to the averaging process. These dynamical equations have the form of an anisotropic generalization to the Friedmann equations obtained for the isotropic solution and reduce to them when isotropy of the scale factor is restored. Unlike the simple case of the isotropic solution, it is unclear how to relate these new terms to a spatial curvature. We use for comparison to observations the isotropic solution. 

We then compare the macroscopic gravity expansion equations to available data sets from distances to supernovae, Baryon Acoustic Oscillations, CMB last scattering surface data, and Hubble constant measurements. We note that instead of using directly $\Omega_A$ in the analysis, we use a term that takes into account its degeneracy with spatial curvature into the Friedmann equation so $\Omega_A$ become a derived parameter.  

We find for the isotropic macroscopic FLRW metric solution $-0.05 \le \Omega_{\mathcal{A}}  \le 0.07$ (at the 95\% confidence level). In the flat metric case, the bounds reduce to $-0.03 \le \Omega_{\mathcal{A}}  \le 0.05$. If we take into account a mathematical and physical prior that restricts the sign of the averaging term to be negative, then the positive part of the interval can be rejected leading to tighter constraints. It is worth noting that the other cosmological parameters ($\Omega_{\Lambda}$, $\Omega_{m}$, $\Omega_{K}$, and $H_{0}$) are moved by a few to several percent from their $\Lambda$CDM concordance model values when the averaging term is included in the analysis. 

Next, we explore the effect of the averaging term on the growth rate of large scale structure. We rederive previous results from perturbing the Macroscopic Gravity field equations and then derive a growth rate equation that can be compared to future observations. We assess the effect of the amplitude of the resulting averaging term on the growth rate function and find that an $\Omega_\mathcal{A}$ term of of amplitude range interval [-0.04,-0.02] lead to an enhancement deviation of the growth up to 2-4\% at late times. This change in the growth is comparable in amplitude to the changes that will be caused by a similar change in the dark energy density parameter or its equation of state. Particularly, the effect of increasing the magnitude of the negative averaging term is to enhance the growth rate of large scale structure which is physically consistent with other results in the literature studying gravitational infall/clustering using exact inhomogeneous cosmological models \cite{infall}. The effect of the averaging on the growth is also comparable in amplitude to some systematic affects in ongoing and future surveys. 

We conclude from using the averaging macroscopic gravity formalism to assess the effect of inhomogeneities on the  expansion history and the growth rate of structure that this effect needs to be tightly constrained and analyzed in the future for a precise and accurate cosmology.

\acknowledgments
We thank J. Dossett for useful comments on modifications to CAMB and CosmoMC. MI acknowledges that this material is based upon work supported in part by NASA under grant NNX09AJ55G and an award from the John Templeton Foundation. Part of the calculations for this work have been performed on the Cosmology Computer Cluster funded by the Hoblitzelle Foundation.

%

\begin{thebibliography}{}
%
\bibitem{Zalaletdinov:1992} R.~Zalaletdinov, Gen. Rel. Grav. {\bf 24} (1992), 1015-103.
%
\bibitem{Zalaletdinov:1993} R.~Zalaletdinov, Gen. Rel Grav. {\bf 25} (1993), 673-695.
%
\bibitem{Behrend} J.~Behrend, (2008), PhD Thesis, [arXiv:0812.2859 [gr-qc]].
%
\bibitem{Boersma} J.~Boersma, \phrd {\bf 57}, (1998), 798-810.
%
\bibitem{Brannlund:2010rs} J.~Brannlund, R.~v.~d.~Hoogen and A.~Coley, Int.\ J.\ Mod.\ Phys.\ D {\bf 19}, 1915 (2010) [arXiv:1003.2014 [gr-qc]].
%
\bibitem{Buchert:1997} T.~Buchert and J.~Ehlers, Astron.\ Astrophys. {\bf 320} (1997), 1-7.
%
\bibitem{Buchert:2003} T.~Buchert and M. Carfora, Phys. Rev. Lett. {\bf 90} (2003), 031101 (1-4).
%
%
%
%
\bibitem{Buchert:2001} T.~Buchert,  Gen.\ Rel.\ Grav.\ {\bf 32} (2000), 105-125. 
%
\bibitem{Buchert:2003} T.~Buchert, Gen. Rel. Grav. {\bf 33} (2001), 1381-1405. 
%
%
\bibitem{Carfora:1996} M.~Carfora and M.~Piotrkowska,  \phrd {\bf 52} (1995), 4393-4424.
%
%
%
\bibitem{Coley:2009yz} A.~A.~Coley, Class.\ Quant.\ Grav.\  {\bf 27}, 245017 (2010) [arXiv:0908.4281 [gr-qc]].
%
%
\bibitem{Debbasch:2004} F.~Debbasch, Eur. Phys. J. B, {\bf 37}(2), (2004), 257.
%
%
\bibitem{Futamase:1988} T.~Futamase, Phys. Rev. Lett. {\bf 61} (1988), 2175-2178.
%
\bibitem{Futamase:1996} T.~Futamase, \phrd {\bf 53} (1996), 681-689.
%
\bibitem{Gasperini:2011} M.~Gasperini, G.~Marozzi, F.~Nugier and G.~Veneziano, J. Cosmol. Astropart. Phys. {\bf 07} (2011), 008.
%
\bibitem{Isaacson:1968f} R.~Isaacson,  Phys. Rev. {\bf 166} (1968), 1263-1271.
%
\bibitem{Isaacson:1968s} R.~Isaacson,  Phys. Rev. {\bf 166} (1968), 1271-1280.
%
\bibitem{Kasai:1993} M.~Kasai,  \phrd {\bf 47} (1993), 3214-3221.
%
\bibitem{Korzynsk:2010} M.~Korzynski, Class. Quantum Grav. {\bf 27} (2010), 105015 (21pp).
%
\bibitem{Noonan:1984} T.~Noonan, Gen. Rel. Grav {\bf 16} 11 (1984), 1103-1118.
%
\bibitem{Noonan:1985} T.~Noonan, Gen. Rel. Grav {\bf 17} 6 (1985), 535-544.
%
\bibitem{Paranjape:2007} A.~Paranjape and T.~Singh, \phrd {\bf 76} (2007), 044006. 
%
\bibitem{Paranjape:2008} A.~Paranjape, \phrd {\bf 78} (2008), 063522.
%
%
\bibitem{Shirokov:1963} M.~Shirokov and I.~Fisher, Sov. Astron. J. {\bf 6} (1963), 699-705.
%
%
%
\bibitem{Zotovv:1992} N.~Zoltov and R.~Stoeger Class. Quantum Grav. {\bf 9} (1992), 1023-1031. 
%
\bibitem{Zotov} N.~Zoltov and R.~Stoeger, Astrophys. J. {\bf 453} 574.
%
\bibitem[Baumann et al.(2012)]{Baumann:2010tm} Baumann, D., Nicolis, 
A., Senatore, L., \& Zaldarriaga, M.\ 2012, \jcap, 7, 051  [arXiv:1004.2488 [astro-ph.CO]].
%
\bibitem{Abramo}
  L.~R.~W.~Abramo, R.~H.~Brandenberger and V.~F.~Mukhanov,
  Phys.\ Rev.\ D {\bf 56}, 3248 (1997)
  [gr-qc/9704037].
%
\bibitem{Andrianomena:2014sya} 
  S.~Andrianomena, C.~Clarkson, P.~Patel, O.~Umeh and J.~P.~Uzan,
  JCAP {\bf 1406}, 023 (2014)
  [arXiv:1402.4350 [gr-qc]].
%
\bibitem[Bagheri 
\& Schwarz(2014)]{2014JCAP...10..073B} Bagheri, S., \& Schwarz, D.~J.\ 2014, \jcap, 10, 073 
\bibitem{Behrend:2007mf} 
  J.~Behrend, I.~A.~Brown and G.~Robbers,
  JCAP {\bf 0801}, 013 (2008)
  [arXiv:0710.4964 [astro-ph]].
%
%
\bibitem{Bertacca} D.~Bertacca, R.~Maartens and C.~Clarckson, (2014), astro-ph.CO 1405.4403
%
\bibitem{Bildhauer:1991} S.~Bildhauer and T.~Futamase, Gen. Rel. Grav. {\bf 23} (1991), 1251-1264 
%
\bibitem{Boehm:2013qqa} 
  C.~Boehm and S.~Rasanen,
  JCAP {\bf 1309}, 003 (2013)
  [arXiv:1305.7139 [astro-ph.CO]].
%
\bibitem{Bolejko:2010wc} K.~Bolejko and R.~A.~Sussman, Phys.\ Lett.\ B {\bf 697}, 265 (2011) [arXiv:1008.3420 [astro-ph.CO]].
%
%
\bibitem{Brown:2013usa} 
  I.~A.~Brown, A.~A.~Coley, D.~L.~Herman and J.~Latta,
  Phys.\ Rev.\ D {\bf 88}, 083523 (2013)
  [arXiv:1308.5072 [gr-qc]].
%
\bibitem{Brown:2012fx} 
  I.~A.~Brown, J.~Latta and A.~Coley,
  Phys.\ Rev.\ D {\bf 87}, no. 4, 043518 (2013)
  [arXiv:1211.0802 [gr-qc]].
%
%
%
%
%
%
%
\bibitem{Bull:2012zx} 
  P.~Bull and T.~Clifton,
  Phys.\ Rev.\ D {\bf 85}, 103512 (2012)
  [arXiv:1203.4479 [astro-ph.CO]].
%
%
\bibitem{Clarkson:2011uk} 
  C.~Clarkson and O.~Umeh,
  Class.\ Quant.\ Grav.\  {\bf 28}, 164010 (2011)
  [arXiv:1105.1886 [astro-ph.CO]].
%
%
%
%
\bibitem{Clarkson:2011br} 
  C.~Clarkson, G.~F.~R.~Ellis, A.~Faltenbacher, R.~Maartens, O.~Umeh and J.~P.~Uzan,
  Mon.\ Not.\ Roy.\ Astron.\ Soc.\  {\bf 426}, 1121 (2012)
  [arXiv:1109.2484 [astro-ph.CO]].
%
\bibitem{Clarkson:2011gm} 
  C.~Clarkson, T.~Clifton, A.~Coley and R.~Sung,
  Phys.\ Rev.\ D {\bf 85}, 043506 (2012)
  [arXiv:1111.2214 [astro-ph.CO]].
%
%
\bibitem[Clifton et al.(2013)]{2013JCAP...11..010C} Clifton, T., Gregoris, 
D., Rosquist, K., \& Tavakol, R.\ 2013, \jcap, 11, 010 
%
%
\bibitem[Di Dio et al.(2012)]{2012JCAP...02..036D} Di Dio, E., Vonlanthen, M., \& Durrer, R.\ 2012, \jcap, 2, 036 %

\bibitem{Ellis:1984} G. F. R.~Ellis,  General Relativity and Gravitation, ed. B Bertotti, F de Felice and A Pascolini (Dordrecht: Reidel, 1984), 215-288
%
\bibitem{Enqvist:2007} K. T.~Enqvist and T.~Mattsson, JCAP 02 (2007) 019.
%
\bibitem{Flanagan:2005dk} 
  E.~E.~Flanagan,
  Phys.\ Rev.\ D {\bf 71}, 103521 (2005)
  [hep-th/0503202].
%
\bibitem{Geshnizjani:2005ce} 
  G.~Geshnizjani, D.~J.~H.~Chung and N.~Afshordi,
  Phys.\ Rev.\ D {\bf 72}, 023517 (2005)
  [astro-ph/0503553].
%
\bibitem{Green:2011} S. R.~Green and R. M. Wald,  \phrd {\bf 83} (2011), 084020 (1-27). 
%
%
%
\bibitem{Hellaby} C.~Hellaby,  Gen. Rel. Grav. 20 (1988), 1203-1217 
%
\bibitem{Hirata:2005ei} 
  C.~M.~Hirata and U.~Seljak,
  Phys.\ Rev.\ D {\bf 72}, 083501 (2005)
  [astro-ph/0503582].
%
%
\bibitem{Ishibashi} A. Ishibashi and R. M.~Wald,  Class. Quantum Grav. 23 (2006), 235-250.
%
%
\bibitem{Kasai:2007} M.~Kasai, Prog.Theor.Phys. {\bf 117} (2007) 1067-1075.
%
%
%
\bibitem{Kolb:2006}  E.~W.~Kolb, S.~Matarrese and A.~Riotto,  New J. Phys. {\bf 8} (2006), 322 (1-25)
%
\bibitem{Larena:2009a} J.~Larena,  \phrd {\bf 79} (2009), 084006 (1-6)
%
\bibitem{Larena:2009b} J.~Larena, J.~M.~Alimi, T.~Buchert, M.~Kunz and P.~S.~Corasaniti, Phys. Rev. D {\bf 79} (2009), 083011 (1-15). 
%
\bibitem{Lavinto:2013exa} 
  M.~Lavinto, S.~RŠsŠnen and S.~J.~Szybka,
  JCAP {\bf 1312}, 051 (2013)
  [arXiv:1308.6731 [astro-ph.CO]].
%
%
\bibitem{Li:2007ny} 
  N.~Li and D.~J.~Schwarz,
  Phys.\ Rev.\ D {\bf 78}, 083531 (2008)
  [arXiv:0710.5073 [astro-ph]].
%
%
\bibitem{Marra:2008} V.~E.~Marra, E.~W.~Kolb and S.~Matarrese,  Phys. Rev. D 77 (2008), 023003 (1-13).
%
\bibitem{Martineau:2005zu} 
  P.~Martineau and R.~Brandenberger,
  astro-ph/0510523.
%
\bibitem{Mattsson:2007qp} 
  T.~Mattsson and M.~Mattsson,
  JCAP {\bf 0802}, 004 (2008)
  [arXiv:0708.3673 [astro-ph]].
%
%
\bibitem{Montani:2003vm} 
  G.~Montani, R.~Ruffini and R.~Zalaletdinov,
  Class.\ Quant.\ Grav.\  {\bf 20}, 4195 (2003)
  [gr-qc/0307077].
%
%
\bibitem{Nambu} Y.~Nambu and M.~Tanimoto, (2005), gr-qc 0507057v1
%
\bibitem{Nugier} F.~Nugier, (2013), arXiv:1309.6542v1
%
\bibitem{Notari} A.~Notari, (2005), astro-ph/0503715
%
%
%
%
\bibitem{Obinna:2011} U.~Obinna, J.~Larena and C.~Clarckson, JCAP 1103 (2011) 029 
%
%
%
%
%
\bibitem{paranjape}
A.~Paranjape and T.~P.~Singh,
  Phys.\ Rev.\ Lett.\  {\bf 101}, 181101 (2008)
  [arXiv:0806.3497 [astro-ph]].
%
\bibitem{Premadi:2001ez} 
  P.~Premadi, H.~Martel, R.~Matzner and T.~Futamase,
  Astrophys.\ J.\ Suppl.\  {\bf 135}, 7 (2001)
  [astro-ph/0101359].
%
%
%
%
%
%
%
\bibitem{Rasanen:2008it} 
  S.~Rasanen,
  JCAP {\bf 0804}, 026 (2008)
  [arXiv:0801.2692 [astro-ph]].
%
\bibitem{Rasanen:2008jcai} S.~R{\"a}s{\"a}nen,  J. Cosmol. Astropart. Phys. {\bf 03} (2010), 018.
%
\bibitem[R{\"a}s{\"a}nen(2012)]{2012PhRvD..85h3528R} S.~R{\"a}s{\"a}nen, 
2012, \prd, 85, 083528 
%
\bibitem{Reiris:2008} M.~Reiris, Class. Quantum Grav. {\bf 25} (2008), 085001 (26pp).
%

 \bibitem{Weinberg} S. Weinberg, Astrophys. J. \textbf{208}, L1 (1976).


\bibitem{Russ:1996km} 
  H.~Russ, M.~H.~Soffel, M.~Kasai and G.~Borner,
  Phys.\ Rev.\ D {\bf 56}, 2044 (1997)
  [astro-ph/9612218].
%
%
\bibitem{Schwarz:2002} J.~D.~Schwarz, 	arXiv:1003.3026 [astro-ph.CO]
%
\bibitem{Seljak} U.~Seljak and L.~Hui ASP Conference Series vol 88, p 267
%
\bibitem{Serui} M.~Seriu, Class. Quantum Grav. {\bf 18} (2001), 5329-5352.
%
\bibitem{Siegel:2005xu} 
  E.~R.~Siegel and J.~N.~Fry,
  Astrophys.\ J.\  {\bf 628}, L1 (2005)
  [astro-ph/0504421].
%
%
\bibitem{Sussman:2011na} 
R.~A.~Sussman,
Class.\ Quant.\ Grav.\  {\bf 28}, 235002 (2011)
[arXiv:1102.2663 [gr-qc]].
%
%
%
%
%
\bibitem{Tanaka:2006us} 
  H.~Tanaka and T.~Futamase,
  Prog.\ Theor.\ Phys.\  {\bf 117}, 183 (2007)
  [astro-ph/0612151].
%
\bibitem{Tanimoto:1999} M.~Tanimoto,  Prog. Theor. Phys. 102 (1999), 1001. 
%
\bibitem{Wetterich:2001kr} 
  C.~Wetterich,
  Phys.\ Rev.\ D {\bf 67}, 043513 (2003)
  [astro-ph/0111166].
%
\bibitem{Wiegand:2010} A.~Wiegand and T.~Buchert, Phys. Rev. D {\bf 82} (2010), 023523 (1-24).
%
\bibitem{Wiltshire:2007a} D.~L.~Wiltshire, New J. Phys. {\bf 9} (2007), 377 (1-66). 
%
%
%


%




 
  %
  %
  %
 %
%
\bibitem{Mars:1997jy}
  M.~Mars and R.~M.~Zalaletdinov,
  J.\ Math.\ Phys.\  {\bf 38} (1997) 4741
  [dg-ga/9703002];
 %

\bibitem{Coley:2005ei}
  A.~A.~Coley, N.~Pelavas and R.~M.~Zalaletdinov,
  Phys.\ Rev.\ Lett.\  {\bf 95} (2005) 151102
  [gr-qc/0504115].
 %
 \bibitem{vandenHoogen:2009nh}
  R.~J.~van den Hoogen,
  J.\ Math.\ Phys.\  {\bf 50} (2009) 082503
  [arXiv:0909.0070 [gr-qc]].
  %
\bibitem{Clifton:2012fs} T.~Clifton, A.~Coley and R.~V.~D.~Hoogen, JCAP {\bf 1210}, 044 (2012) [arXiv:1209.1085 [astro-ph.CO]].
  
 %
  \bibitem{J.L.Synge:1960zz}
  J.~L.~Synge,
  North-Holland, Amsterdam, 1960
    %
 
 %
 %


 \bibitem[Weinberg(1976)]{1976ApJ...208L...1W} Weinberg, S.\ 1976, \apjl, 208, L1 
\bibitem{Suzuki:2011hu} 
  N.~Suzuki, D.~Rubin, C.~Lidman, G.~Aldering, R.~Amanullah, K.~Barbary, L.~F.~Barrientos and J.~Botyanszki {\it et al.},
  Astrophys.\ J.\  {\bf 746}, 85 (2012)
  [arXiv:1105.3470 [astro-ph.CO]].
\bibitem{Hinshaw:2012aka}
  G.~Hinshaw {\it et al.}  [WMAP Collaboration],
  Astrophys.\ J.\ Suppl.\  {\bf 208} (2013) 19
  [arXiv:1212.5226 [astro-ph.CO]].
  %
  \bibitem{Blake:2011en}
  C.~Blake, E.~Kazin, F.~Beutler, T.~Davis, D.~Parkinson, S.~Brough, M.~Colless and C.~Contreras {\it et al.},
  Mon.\ Not.\ Roy.\ Astron.\ Soc.\  {\bf 418} (2011) 1707
  [arXiv:1108.2635 [astro-ph.CO]].
  %
  %
  \bibitem{Riess:2009pu}
  A.~G.~Riess, L.~Macri, S.~Casertano, M.~Sosey, H.~Lampeitl, H.~C.~Ferguson, A.~V.~Filippenko and S.~W.~Jha {\it et al.},
  Astrophys.\ J.\  {\bf 699} (2009) 539
  [arXiv:0905.0695 [astro-ph.CO]].
  %
  \bibitem{Lewis:2002ah}
  A.~Lewis and S.~Bridle,
  Phys.\ Rev.\ D {\bf 66} (2002) 103511
  [astro-ph/0205436].
  %
  %
\bibitem{Komatsu:2010fb} 
  E.~Komatsu {\it et al.}  [WMAP Collaboration],
  Astrophys.\ J.\ Suppl.\  {\bf 192}, 18 (2011)
  [arXiv:1001.4538 [astro-ph.CO]].
\bibitem{Hu:1995en} 
  W.~Hu and N.~Sugiyama,
  Astrophys.\ J.\  {\bf 471}, 542 (1996)
  [astro-ph/9510117].
\bibitem{Eisenstein:2005su}
  D.~J.~Eisenstein {\it et al.}  [SDSS Collaboration],
  Astrophys.\ J.\  {\bf 633} (2005) 560
  [astro-ph/0501171].
  \bibitem{Peel:2012vg}
  A.~Peel, M.~Ishak and M.~A.~Troxel,
  Phys.\ Rev.\ D {\bf 86} (2012) 123508
  [arXiv:1212.2298 [astro-ph.CO]].
%
  \bibitem{Bolejko}
  K. Bolejko, Phys.\ Rev.\ D {\bf 75} (2007) 043508.
  

\bibitem{Kessler:2009ys} 
  R.~Kessler, A.~Becker, D.~Cinabro, J.~Vanderplas, J.~A.~Frieman, J.~Marriner, T.~MDavis and B.~Dilday {\it et al.},
  Astrophys.\ J.\ Suppl.\  {\bf 185}, 32 (2009)
  [arXiv:0908.4274 [astro-ph.CO]].
\bibitem{Wald:1984rg} 
  R.~M.~Wald,
  Chicago, Usa: Univ. Pr. (1984) 491p
 \bibitem{kodamaetall1984} H. Kodama, M. Sasaki, Prof. of. Theor. Phys. Sup. 78 (1984)
 
 \bibitem{19} L. Guzzo \textit{et al.}, Nature, \textbf{451}, 541 (2008). 
\bibitem{20} M. Colless \textit{et al.}, Mon. Not. R. Astron. Soc. \textbf{328}, 1039 (2001). 
\bibitem{27} C. Blake \textit{et al.}, Mon. Not. R. Astron. Soc. \textbf{415}, 2876 (2011).
\bibitem{21} M. Tegmark \textit{et al.}, Phys. Rev. D \textbf{74}, 123507 (2006). 
\bibitem{22} N. P. Ross \textit{et al.}, Mon. Not. R. Astron. Soc. \textbf{381}, 573 (2007). 
\bibitem{23} J. da \^Angela \textit{et al.}, Mon. Not. R. Astron. Soc. \textbf{383}, 565 (2008). 
\bibitem{25} M. Viel, M. G. Haehnelt, and V. Springel, Mon. Not. R. Astron. Soc. \textbf{354}, 684 (2004). 
\bibitem{26} M. Viel and M. G. Haehnelt, Mon. Not. R. Astron. Soc. \textbf{365}, 231 (2006).
\bibitem{24} P. McDonald \textit{et al.}, Astrophys. J. \textbf{635}, 761 (2005).
 
 
 \bibitem{dossett} J. Dossett, M Ishak, Y Gong and A. Wang, J. Cosmol. Astropart. Phys. 04 (2010) 022.
\bibitem{linder} E.V. Linder, Phys. Rev. D {\bf 72}, 043529 (2005).
\bibitem{polarski} D. Polarski and R. Gannouji, Phys. Lett. B {\bf 660}, 439 (2008).
\bibitem{Gong09}  Y. Gong, M. Ishak, A. Wang, arXiv:0903.0001v1 [astro-ph.CO] (2009).
\bibitem{Mortonson} M. J. Mortonson, W. Hu and D. Huterer, Phys. Rev. D {\bf 79}, 023004 (2009).
\bibitem{linder07} E.V. Linder and R.N. Cahn, Astropart. Phys. {\bf 28}, 481 (2007).
\bibitem{gannouji} R. Gannouji and D. Polarski, J. Cosmol. Astropart. Phys. 05 (2008) 018.
\bibitem{ygong} Y. Gong, M. Ishak, and A. Wang, Phys. Rev. D \textbf{80}, 023002 (2009).
\bibitem{PaperI} M. Ishak, J. Dossett, Phys. Rev. D {\bf 80}, 043004 (2009)




%
  \bibitem{infall} M. A. Troxel, A. Peel, M. Ishak, J. Cosmol. Astropart. Phys. 12 (2013) 048.
  \bibitem{VanDenHoogen:2007en} 
  R.~J.~Van Den Hoogen,
  Gen.\ Rel.\ Grav.\  {\bf 40}, 2213 (2008)
  [arXiv:0710.1823 [gr-qc]].
\bibitem{Coley:2006} 
  A.~A.~Coley and N.~Pelavas,
  Phys.\ Rev.\ D {\bf 74}, 087301 (2006)
  [astro-ph/0606535].
  A.~A.~Coley and N.~Pelavas,
  Phys.\ Rev.\ D {\bf 75}, 043506 (2007)
  [gr-qc/0607079].
%
 \bibitem{mix} 
  H.~Stephani, E.~Herlt, M.~MacCullum, C.~Hoenselaers and D.~Kramer,
  ``Exact Solutions of Einstein's Equations,'';
  L.~D.~Landau and E.~M.~Lifschits,
  ``The Classical Theory of Fields : Course of Theoretical Physics, Volume 2,''.
 \bibitem{Misner:1967zz} 
  C.~W.~Misner,
  Phys.\ Rev.\ Lett.\  {\bf 19}, 533 (1967).
%
\end{thebibliography}
\end{document}